\documentclass[reprint,onecolumn,notitlepage,tightenlines,nofootinbib,eqsecnum,superscriptaddress,preprintnumbers,10pt]{revtex4-1}
\pdfoutput=1
\usepackage{epsfig,amsfonts,mathrsfs,amsmath,amssymb,graphicx,color,slashed}
\usepackage{amsmath,latexsym,amssymb,graphicx,slashed,hyperref,color,enumerate,url,etoolbox}
\hypersetup{colorlinks,citecolor= nicegreen,linkcolor= nicered}
\definecolor{nicered}{rgb}{0.7,0.1,0.1}
\definecolor{nicegreen}{rgb}{0.1,0.5,0.1}

\newcommand{\tcg}[1]{\textcolor{nicegreen}{#1}}
\def\LeNCS{\sl LeNCS}
\def\BmL{$B$$-$$L$}

\def\VT{Center for Neutrino Physics, Department of Physics, Virginia Tech, Blacksburg, VA 24061, USA}
\def\Northwestern{Department of Physics and Astronomy, Northwestern University, Evanston, IL 60208, USA}
\def\Fermilab{Theoretical Physics Department, Fermilab, P.O. Box 500, Batavia, IL 60510, USA}

\begin{document}

\allowdisplaybreaks

%%%%%%%%%%%%%%%%%%%%%%%%%%%%%%%%%%%%%%%%%%
%%%%%%%%%%                            New Title                     %%%%%%%%%%%%
%%%%%%%%%%%%%%%%%%%%%%%%%%%%%%%%%%%%%%%%%%

\title{Lepton-Number-Charged Scalars and Neutrino Beamstrahlung}

\author{Jeffrey M. Berryman}
\affiliation{\VT}
\author{Andr\'e de Gouv\^ea}
\affiliation{\Northwestern}
\author{Kevin J. Kelly}
\affiliation{\Northwestern}
\author{Yue Zhang}
\affiliation{\Northwestern}
\affiliation{\Fermilab}
\date{\today}

\begin{abstract}
Experimentally, baryon number minus lepton number, {\BmL}, appears to be a good global symmetry of nature. We explore the consequences of the existence of gauge-singlet scalar fields charged under {\BmL} -- dubbed lepton-number-charged scalars, {\LeNCS} -- and postulate that these couple to the standard model degrees of freedom in such a way that {\BmL} is conserved even at the non-renormalizable level. In this framework, neutrinos are Dirac fermions. Including only the lowest mass-dimension effective operators, some of the {\LeNCS} couple predominantly to neutrinos and may be produced in terrestrial neutrino experiments. We examine several existing constraints from particle physics, astrophysics, and cosmology to the existence of a {\LeNCS} carrying {\BmL} charge equal to two, and discuss the emission of {\LeNCS}'s via ``neutrino beamstrahlung,'' which occurs every once in a while when neutrinos scatter off of ordinary matter. We identify regions of the parameter space where existing and future neutrino experiments, including the  Deep Underground Neutrino Experiment, are at the frontier of searches for such new phenomena. 
\end{abstract}
\preprint{NUHEP-TH/18-03, FERMILAB-PUB-18-020-T}
\maketitle

\pagestyle{plain}

%%%%%%%%%%%%%%%%%%%%%%%%%%%%%%%%%%%%%%%%%%
%%%%%%%%%%                       New Section I                   %%%%%%%%%%%%
%%%%%%%%%%%%%%%%%%%%%%%%%%%%%%%%%%%%%%%%%%

\section{Introduction}

The gauge symmetry and particle content of the Standard Model (SM) are such that, at the renormalizable level, both $U(1)_B$ -- baryon number -- and $U(1)_L$ -- lepton number -- are exact classical global symmetries of the Lagrangian. Both turn out to be anomalous and hence violated at the quantum level. The combined $U(1)_{B-L}$ -- baryon number minus lepton number ({\BmL}) -- however, turns out to be anomaly-free and is an excellent candidate for a fundamental symmetry of nature, black-hole arguments notwithstanding (see, e.g., Refs.~\cite{Abbott:1989jw,Coleman:1993zz,Kallosh:1995hi} and many references therein). Experimentally, there is no evidence for the non-conservation of either baryon number or lepton number, in spite of ambitious, ultra-sensitive, decades-long experimental enterprises~\cite{Patrignani:2016xqp,Babu:2013jba}. 

If $U(1)_{B-L}$ is a fundamental symmetry of nature, nonzero neutrino masses require the existence of new fermions charged under $U(1)_{B-L}$. We choose these to be left-handed antineutrinos $\nu^c$ (the conjugated states to the right-handed neutrinos, to use a more familiar name), with lepton number $-1$ and {\BmL} charge +1. Experiments require the existence of at least two flavors of $\nu^c$ fields and, unless otherwise noted, we assume there are three. Conserved {\BmL} implies that neutrinos are Dirac fermions and nonzero masses are a consequence of neutrino Yukawa interactions, 
\begin{equation}
\mathcal{L}_{\text{Yuk}} \supset y_\nu L H \nu^c + {\rm h.c.}, \, 
\end{equation}
where $H$ is the Higgs doublet (hypercharge $+1/2$), $L$ are the lepton doublets, and $y_{\nu}$ are the neutrino Yukawa couplings. Flavor indices are suppressed. After electroweak symmetry breaking, the neutrino Dirac mass matrix is $m_{\nu}=y_{\nu}v/\sqrt{2}$, where $v=246$~GeV and $v/\sqrt{2}$ is the vacuum expectation value (vev) of the neutral component of the Higgs field. Experimental constraints require $y_{\nu}$ to be of order $10^{-12}$ or smaller~\cite{Patrignani:2016xqp,deGouvea:2013onf}.

Here we are interested in the consequences of allowing for the existence of new degrees of freedom charged under  $U(1)_{B-L}$, assuming {\BmL} is conserved even if one allows for  higher-dimensional operators. More specifically, we explore the physics of new scalar fields with nonzero {\BmL} charge but which are otherwise singlets of the SM gauge interactions. We will refer to these as lepton-number-charged scalars, or {\LeNCS}. Different combinations of {\LeNCS} fields have different non-trivial {\BmL} charge and their couplings to the SM will be guided by {\BmL} conservation. Since we treat {\BmL} as an exact symmetry, we assume the scalar potential is such that none of the {\LeNCS} fields acquire vevs.

The field content of the SM, augmented to include the $\nu^c$ fields, is such that all gauge-invariant, Lorentz-invariant operators have even {\BmL} charges. Furthermore, it has been shown \cite{Rao:1983sd,deGouvea:2014lva,Kobach:2016ami} that, for any operator of mass-dimension $d$ and $B-L$ charge $q_{B-L}$,  given the SM field content plus the $\nu^c$ fields, 
\begin{equation}
(-1)^d=(-1)^{q_{B-L}/2} \,. 
\end{equation}
As advertised, since $d$ are integers, the {\BmL} charge of any operator is even. Odd-dimensional operators have {\BmL} charge $4n+2$ while even-dimensional ones have {\BmL} charge $4n$, where $n$ is an integer. This automatically implies that all {\LeNCS} species with odd {\BmL} charge can only couple to the SM fields in pairs, while it is possible for {\LeNCS} species with even {\BmL} charge to couple individually to the SM.

In this paper, we will concentrate on a {\LeNCS} with {\BmL} charge equal to +2, denoted hereafter by $\phi$. A single $\phi$ can couple to {\BmL}-charge-two gauge-invariant SM operators. As discussed above, these are odd-dimensional, and the lowest-order ones are $\nu^c \nu^c$ (dimension three), and $(L H)(L H)$ (dimension five). Hence, up to dimension six, the most general Lagrangian that describes the SM augmented by the $\phi$-field, assuming $U(1)_{B-L}$ is a good symmetry of nature, includes
\begin{equation}\label{EFToperators5}
\mathcal{L}_{\rm \phi} \supset \frac{\lambda^{ij}_c}{2} \nu^c_i \nu^c_j \phi^* +  \frac{(L_{\alpha} H)(L_{\beta} H)}{\Lambda^2_{\alpha\beta}} \phi  \\
+ {\rm h.c.}\, .
\end{equation}
Here, $\lambda_c^{ij}$, $i,j=1,2,3$ (neutrino mass-eigenstate labels) are dimensionless couplings while $\Lambda^2_{\alpha\beta}$ are dimensionful couplings, $\alpha,\beta=e,\mu,\tau$ (lepton flavor labels). We have ignored operators that contain derivatives. In Appendix~\ref{app:dim8}, we list the operators with mass-dimension eight that contain a single $\phi$ field. Up to dimension five, there is only one interaction term that contains two $\phi$ fields, $|H|^2|\phi|^2$. The scalar potential for $\phi$, therefore, contains
\begin{equation}\label{scalarpotential}
\mathcal{L}_{\rm |\phi|^2} = -{\mu}_{\phi}^2|\phi|^2 - c_{\phi}|\phi|^4 - c_{\phi H}|\phi|^2|H|^2, 
\end{equation}
where $\mu_{\phi}$ is the $\phi$ mass-squared parameter, while $c_{\phi},c_{\phi H}$ are dimensionless quartic couplings. Excluding the effects of nonrenormalizable operators, the mass-squared of the $\phi$ field is $m_{\phi}^2 = {\mu}_{\phi}^2 + c_{\phi H}v^2/2$.

We will concentrate on the consequences of Eq.~\eqref{EFToperators5} assuming that the $\phi$ mass is smaller than the electroweak symmetry breaking scale. We will be especially interested in the $\Lambda^2_{\alpha\beta}$ couplings but will also discuss consequences of $\lambda_c$. Throughout, we will assume that the $c_{\phi}$ and $c_{\phi H}$ couplings are small enough that they do not lead to any observable consequences in the laboratory or in the early universe. We will briefly return to potential apparent baryon-number-violating effects of higher-dimensional operators later. 

After electroweak symmetry breaking, Eq.~\eqref{EFToperators5} becomes
\begin{equation}\label{Lint}
\mathcal{L}_{\rm int} = \frac{\lambda_c^{ij}}{2} \nu^c_i \nu^c_j \phi^* + \frac{\lambda_{\alpha\beta}}{2} \nu_{\alpha} \nu_{\beta}\phi + \frac{\lambda_{\alpha\beta}}{v} \nu_{\alpha} \nu_{\beta} \phi h + {\rm h.c.} + \mathcal{O}(h^2) \ ,
\end{equation}
where $\lambda_{\alpha\beta} =   \lambda_{\beta\alpha} = v^2/\Lambda_{\alpha\beta}^2$ are the elements of a  symmetric matrix of dimensionless couplings. We expand the physical Higgs boson field $h$ around its vacuum expectation value $v$ up to linear order. In this simple setup, the new scalar $\phi$ couples predominantly to SM neutrinos, and we will demonstrate that it could lead to interesting, observable effects in neutrino experiments. For example, $\phi$ could be radiated when neutrinos scatter off of regular matter. Because $\phi$ carries away lepton number, the neutrino charged-current interaction will lead to wrong-sign charged leptons. This $\phi$ radiation would not only change the charged-lepton energy spectrum but also lead to significant missing transverse energy in the event.\footnote{As discussed above, {\LeNCS} fields with odd {\BmL} charge have to be emitted in pairs from neutrino beams. The corresponding rate is more phase-space suppressed compared to single-$\phi$ radiation and, as a result, their impact on neutrino experiments is less significant. On the other hand, {\LeNCS} fields with odd {\BmL} charge serve as dark matter candidates. We will elaborate on this to some extent in Section~\ref{sec:DM}.} We explore these effects in a few accelerator neutrino experiments, including NOMAD, MiniBooNE, MINOS, NO$\nu$A, and DUNE. For $\phi$ to be efficiently radiated in these experiments, we are interested in $m_\phi$ values around or below the GeV scale and $\lambda \sim \mathcal{O}(1)$.
In contrast, the $\lambda_c$ couplings between $\phi$ and the right-handed neutrinos are mostly inconsequential to laboratory experiments. On the other hand, in conjunction with the $\lambda$ or the $c_{\phi H}$ couplings, $\lambda_c$ effects can leave an imprint in cosmological observables, including the number of relativistic degrees of freedom at the time of big-bang nucleosynthesis~\cite{Cyburt:2004yc, Barger:2003zg} and the formation of the cosmic microwave background (CMB)~\cite{Ade:2015xua}.  

One last point before proceeding. The effective coupling of $\phi$ to neutrinos has a similar form to that of a Majoron~\cite{Chikashige:1980ui, Gelmini:1980re}, the (pseudo) goldstone boson in models where lepton number is broken in a controlled way. The equivalent coupling $\lambda$ for a Majoron is related to the observed neutrino masses, $\lambda \sim m_\nu/f$, where $f$ is the spontaneous lepton number breaking scale. The mass of the Majoron is generated by explicit lepton-number-violating interactions and, in most cases, is smaller than $f$. The theory has an approximate lepton-number symmetry at energy scales above $f$. The new phenomena related to the {\LeNCS} $\phi$ to be discussed in the rest of this work cannot be mimicked by a Majoron. The neutrino couplings to $\phi$ here are in no way related to or constrained by the observed neutrino masses. Moreover, we will be mostly interested in sizable couplings, $\lambda \sim \mathcal{O}(1)$. If this were the coupling of a Majoron, then the lepton-number-breaking scale would have to be very low, $f \sim \mathcal{O}(\rm eV)$; under these circumstances, it would be inconsistent to talk about Majorons with masses above an MeV.

%%%%%%%%%%

\section{Existing constraints}\label{sec:limits}

While the new scalar $\phi$ mainly interacts with neutrinos, there are several indirect constraints on the couplings $\lambda$, defined above in Eq.~\eqref{Lint}. In this section, we derive and discuss these constraints, which will set the stage for the discussions in the next section on the effects of $\phi$ in neutrino experiments. Many of the low-energy constraints discussed here also apply to other models with new neutrino interactions mediated by new light bosons, see for example Refs.~\cite{Ng:2014pca, Cherry:2014xra}. 

\subsection{Invisible Higgs decay}

\begin{figure*}
\centerline{\includegraphics[width=0.9\textwidth]{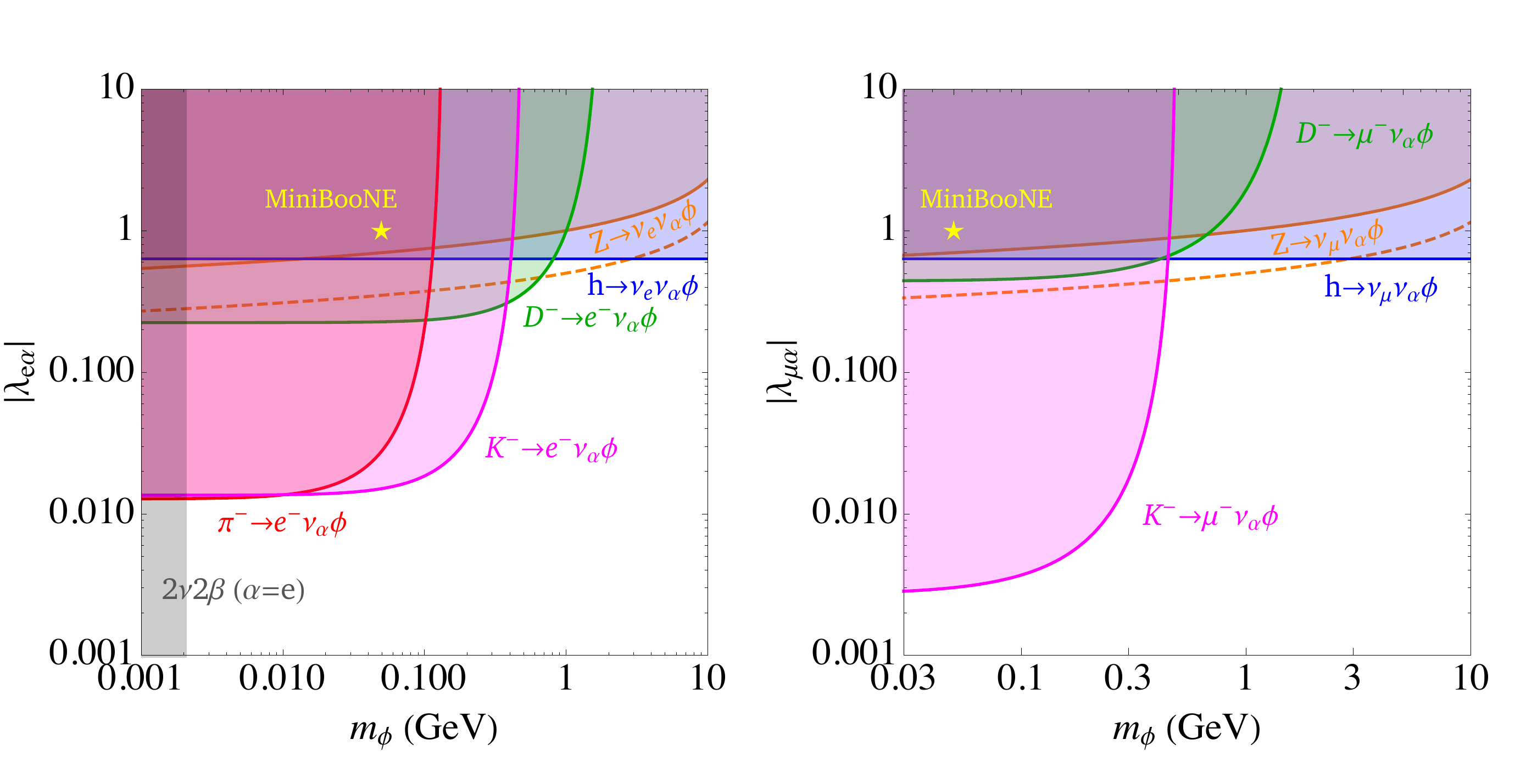} }
\caption{Experimental constraints on the couplings between a {\LeNCS} $\phi$ and neutrinos, $\lambda_{e\alpha}$ (left plot), $\lambda_{\mu\alpha}$ (right plot), with $\alpha=e,\mu,\tau$, as a function of its mass $m_{\phi}$. The colorful regions in the plots are ruled out by precision measurements of the decays of pions (red), $K$-mesons (pink), $D$-mesons (green), $Z$-bosons (orange), and Higgs bosons (blue). In the case of $Z$-boson decays, the lower limit is weaker for flavor diagonal $\lambda$ couplings (solid orange curve) than flavor off-diagonal couplings (dashed orange curve). For the $\lambda_{ee}$ coupling, there is an additional constraint from measurements of double-beta nuclear decay rates.  The yellow stars indicate points in the parameter spaces where one obtains a reasonably good fit to the MiniBooNE low energy excess, as discussed in Sec.~\ref{subsec:miniboone}.}
\label{allconstraints}
\end{figure*}

A light scalar could have an impact on the decay of the 125\,GeV Higgs boson. If $m_\phi<m_h$ the third interaction term in Eq.~\eqref{Lint} allows for a new decay channel for the Higgs boson, $h\to \nu_\alpha \nu_\beta \phi$. The $\phi$ particle in the final state further decays into two antineutrinos, a decay mediated by the first two interaction terms in Eq.~\eqref{Lint}. Hence, nonzero $\lambda$ leads to an invisible decay mode for the Higgs boson. The three-body decay width into a fixed combination of flavors $\alpha,\beta$ is
\begin{equation}
\Gamma(h\to\nu_\alpha \nu_\beta\phi) \simeq \frac{|\lambda_{\alpha\beta}|^2  m_h^3}{384\pi^3 v^2} \ ,
\end{equation}
where the dependency on $m_{\phi}$ has been suppressed since we are mostly interested in $m_\phi\sim {\rm GeV} \ll m_h$. Assuming only one $\lambda_{\alpha\beta}$ is nonzero,  
the current upper limit on the Higgs invisible decay branching ratio, defined as 
\begin{equation}
{\rm Br}(h_{inv}) = \frac{\Gamma(h\to\nu_{\alpha} \nu_{\beta}\phi)+\Gamma(h\to\bar \nu_{\alpha} \bar\nu_{\beta}\phi^*)}{\Gamma(h\to\nu_{\alpha} \nu_{\beta}\phi)+\Gamma(h\to\bar\nu_{\alpha} \bar\nu_{\beta}\phi^*) + \Gamma^h_{\rm SM}} \ ,
\end{equation}
is less than $34\%$~\cite{Khachatryan:2016vau}, and translates into
\begin{equation}
|\lambda_{\alpha\beta}|\lesssim0.7 \ .
\end{equation}
This constraint is depicted in blue in Fig.~\ref{allconstraints} in the $m_{\phi}\times \lambda_{\beta\alpha}$--plane, for $\beta=e,\mu$. Identical constraints apply for $\lambda_{\tau\alpha}$.

It is worthwhile emphasizing that we consider values of $\lambda$ less than or equal to one. These, in turn, imply $\Lambda$ values -- see Eq.~\eqref{EFToperators5} -- of order the weak scale or higher. When using the effective theory to describe processes at the electroweak scale (e.g., $Z$-boson and Higgs boson decays), the effective theory approach may still qualify as a faithful description of the system, especially if one allows the physics responsible for the dimension-six operator in Eq.~\eqref{EFToperators5} to be somewhat strongly coupled. We return to this issue in Section~\ref{sec:UV}.

\subsection{Invisible $Z$-boson decay}

If $m_\phi<M_Z$ there is an additional invisible $Z$-boson decay mode, $Z\to\nu_{\alpha}\nu_{\beta}\phi$ where $\phi$ is radiated from one of the final state neutrinos. This three-body decay can potentially be important for light $\phi$ because of a collinear enhancement of the decay width for the radiation of a light $\phi$ particle. For $m_\phi\ll M_Z$, the decay rate takes the form, for fixed final-state neutrino flavors $\alpha,\beta$,
\begin{equation}
\Gamma(Z\to\nu_{\alpha} \nu_{\beta}\phi) \simeq \frac{G_F M_Z^3 |\lambda_{\alpha\beta}|^2 \left( \ln\frac{M_Z}{m_\phi} - \frac{5}{3} \right)}{288\sqrt{2}\pi^3 (1+\delta_{\alpha\beta})^2} \ ,
\end{equation}
where the factor of $\delta_{\alpha\beta}$ arises for identical final-state neutrinos. The measured $Z$-boson invisible decay branching ratio is $(20\pm0.06)\%$ and the $Z$-boson total width is 2.495\,GeV~\cite{LEP:2003aa}. This translates into
\begin{equation}
|\lambda_{\alpha\beta}|<0.5 (1+\delta_{\alpha\beta}) \ ,
\end{equation}
for $m_\phi=1\,$GeV. This constraint is depicted in orange in Fig.~\ref{allconstraints} in the $m_{\phi}\times \lambda_{\beta\alpha}$--plane, for $\beta=e,\mu$. The solid (dashed) line applies for $\alpha=\beta$ ($\alpha\neq\beta$). Identical constraints apply for $\lambda_{\tau\alpha}$.

\subsection{Charged meson decays}
\label{sec:meson}

If $m_\phi$ is smaller than one GeV there are strong constraints from the decays of charged pseudoscalar mesons $M^\pm = \pi^\pm, K^\pm, D^\pm$ \cite{Pasquini:2015fjv}. In particular, the decay rate associated with $\phi$-emission -- the $\phi$ is radiated from the final state neutrino -- is not proportional to the well known helicity-suppression factor associated with two-body leptonic decays of charged pseudoscalars.

The decay width of $M^-\to \ell_{\alpha}^- \nu_{\beta} \phi$ is
\begin{equation}\label{nohelicitysup}
\Gamma(M^-\to \ell_{\alpha}^- \nu_{\beta} \phi)  = \frac{|\lambda_{\alpha\beta}|^2 G_F^2 f_M^2}{768 \pi^3 m_M^3} 
\left[ (m_M^2 - m_\phi^2)(m_M^4 + 10 m_M^2 m_\phi^2 + m_\phi^4)  \rule{0mm}{5mm} - 12 m_M^2 m_\phi^2 (m_M^2+m_\phi^2) \ln\frac{m_M}{m_\phi} \right] \ .
\end{equation}
We translate experimental measurements of, or constraints on, $M^-\to \ell_{\alpha}^- \bar\nu_{\beta}$ or $\ell_{\alpha}^- \bar\nu_{\beta} \nu\bar\nu$ decays as bounds on the above decay rate. These are summarized in the table below~\cite{Patrignani:2016xqp}. These constraints are depicted in red/pink/green for $\pi/K/D$-mesons in Fig.~\ref{allconstraints} in the $m_{\phi}\times \lambda_{\beta\alpha}$--plane, for $\beta=e,\mu$. More detailed estimates, consistent with the ones listed here, can be found in Ref.~\cite{Pasquini:2015fjv}.

\begin{table}[h]
\centering\begin{tabular}{c|c|c}
\hline
Decay channels from PDG  & Decay channels in our model        &  Upper bound on Br \\  %&  $m_M$/MeV & $f_M$/MeV\\
\hline
$\pi\to e\bar\nu_e\nu\bar\nu$  & $\pi\to e \nu_\alpha \phi$          &  $5\times 10^{-6}$ \\ %& 137 & 131 \\
\hline
$K\to e\bar\nu_e\nu\bar\nu$   &  $K\to e\nu_\alpha \phi$            &  $6\times 10^{-5}$\\%  & $1.2\times10^{-8}$ &  494 & 160 \\
\hline
$K\to \mu\bar\nu_\mu\nu\bar\nu$   &  $K\to \mu\nu_\alpha \phi$      &      $2.4\times 10^{-6}$\\%  & $1.2\times10^{-8}$ &  494 & 160 \\
\hline
$D\to e\bar\nu_e$   &  $D\to e\nu_\alpha \phi$      &      $8.8\times 10^{-6}$\\%  & $1.04\times10^{-12}$ &  1869 & 249 \\
\hline
$D\to \mu\bar\nu_\mu$   &  $D\to \mu\nu_\alpha \phi$      &      $ 3.4\times 10^{-5}$\\%  & $1.04\times10^{-12}$ &  1869 & 249 \\
\hline
\end{tabular}
\end{table}
We also examined constraints from $\tau$ decays and final state $\phi$ radiation, and found limits that are weaker than those estimated above using meson decays. $\tau$ decays, however, also provide information concerning $\lambda_{\tau\tau}$.

\subsection{Double-beta decays}

If $m_\phi$ were smaller than a few MeV, then  $\phi$ could provide new double-beta decay channels for certain nuclei. In particular light $\phi$ particles can be produced via virtual neutrino annihilation, 
\begin{eqnarray}
(Z,A)\to (Z+2,A) e^-e^-\phi \ .
\end{eqnarray}
This is identical to Majoron emission. Recent measurements of double-beta decay rates~\cite{Agostini:2015nwa} translate into an upper bound on 
\begin{eqnarray}
|\lambda_{ee}|\lesssim10^{-4} \ .
\end{eqnarray} 
This constraint is depicted by a grey band in Fig.~\ref{allconstraints} in the $m_{\phi}\times \lambda_{e\alpha}$--plane, keeping in mind it only applies for $\alpha=e$. 

\subsection{Charged-lepton flavor violation}

\begin{figure}[h]
\centerline{\includegraphics[width=0.35\textwidth]{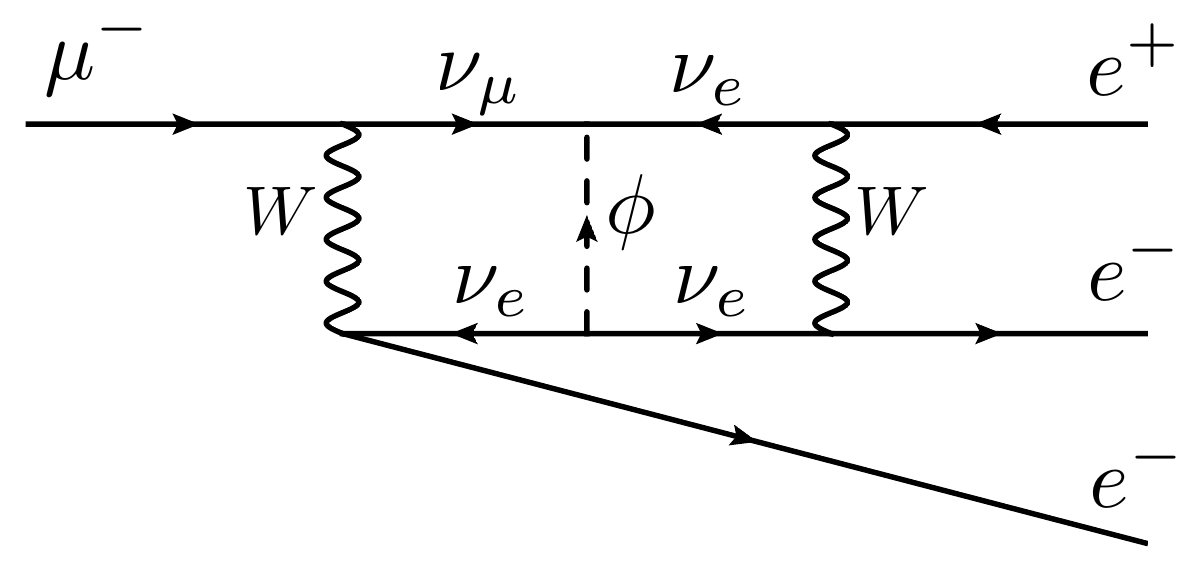} }
\caption{A representative Feynman diagram for $\mu\to3e$, which occurs at two-loop order. The time direction is from left to right.}\label{fig:LFV}
\end{figure}

At loop level, $\phi$-exchange mediates charged-lepton flavor-violating processes. At two loops, for example, the exchange of $\phi$ and two $W$-bosons in a double-box diagram leads to the rare muon decay process $\mu\to3e$. A representative Feynman diagram is depicted in Fig.~\ref{fig:LFV}.

We estimate that the current constraint ${\rm Br}(\mu\to 3e)<10^{-12}$~\cite{Bellgardt:1987du} translates into 
\begin{eqnarray}
|\lambda_{ee} \lambda_{e\mu}| \lesssim 10^{-2} \ .
\end{eqnarray}
This and other similar constraints are not depicted in Fig.~\ref{allconstraints} and will be ignored henceforth because they involve products of two different $\lambda_{\alpha\beta}$ couplings and can be avoided simply by assuming that some of the $\lambda_{\alpha\beta}$ are much smaller than the others.

\subsection{Cosmological constraint}
 
The virtual exchange of $\phi$ will lead to neutrino self-interactions that can be parameterized by a four-fermion operator with an effective coupling constant $G_{\rm eff} \simeq |\lambda|^2/m_\phi^2$, for large enough values of $m_{\phi}$. The strength of this effective interaction is constrained by cosmological observations sensitive to the neutrino free streaming length in the early universe. A recent analysis of the CMB power spectrum data derives an upper bound $G_{\rm eff}<10^8 G_F$~\cite{Oldengott:2017fhy}. This, in turn, translates into 
\begin{eqnarray}
\lambda \lesssim \frac{m_\phi}{30\,{\rm MeV}} \ .
\end{eqnarray}
For GeV-scale $\phi$ mass, this constraint is weaker than those discussed above, see Fig.~\ref{allconstraints}.

\subsection{Supernova 1987A}

The $\phi$ particle, if light enough, could be radiated through its interaction with neutrinos and accelerate the cooling of core-collapse supernovae. 
The impact of a light scalar on the observations of SN1987A is similar to that of a Majoron, studied in Ref.~\cite{Choi:1987sd}, and revisited more recently in Refs.~\cite{Farzan:2002wx, Heurtier:2016otg}. The analysis of Ref.~\cite{Heurtier:2016otg} finds that for 1\,MeV$\lesssim m_\phi\lesssim 100\,$MeV, value of $\lambda_{\alpha\beta}$ between $10^{-12}$ to $10^{-6}$ are excluded, for $\alpha, \beta = e, \mu$. For larger couplings, the $\phi$ particles are reabsorbed by the supernova and no longer affect its cooling. As a result, this bound corresponds to a rather weak coupling, well below the region of the parameter space depicted in Fig.~\ref{allconstraints}.

\subsection{$N_{\rm eff}$ constraint on $\lambda_c$ couplings}
\label{sec:lc}

The couplings $\lambda_c$, defined in the first line of Eq.~\eqref{EFToperators5}, are mostly unconstrained as long as $m_{\phi}$ is heavier than the neutrinos. This is easy to understand: $\lambda_c$ mediates interactions between right-handed neutrinos and the $\phi$ field and, as long as the neutrinos are ultra-relativistic, right-handed neutrino properties are virtually unconstrained. For large enough values of $\lambda_c$ and light enough values of $m_{\phi}$, however, we anticipate $\lambda_c$ effects in neutrino decay and in the dynamics of relic neutrinos. We do not consider these constraints here but hope to return to them in future work. 

Since we are interested in GeV-scale $\phi$ with $\mathcal{O}(1)$ coupling $\lambda$ between the left-handed neutrinos and $\phi$, however, we need to appreciate the fact that, in the early universe, $\phi$-exchange between the right-handed and left-handed neutrino degrees of freedom may lead to a thermal right-handed neutrino population. Within this context, the cosmological observation of $\Delta N_{\rm eff}=-0.01\pm 0.18$~\cite{Ade:2015xua}
constrains the couplings $\lambda_c$. A safe way to evade this constraint is to require the right-handed neutrinos to decouple from the SM plasma at temperatures above the QCD phase transition~\cite{Steigman:2013yua}. This translates into\footnote{There is currently a 3.4$\sigma$ discrepancy between the value of the Hubble constant indicated by local observations of the expansion of the Universe and the value determined by the Planck collaboration. It has been argued that analyzing Planck data jointly with the local measurements translates into $\Delta N_{\rm eff} \lesssim 1$ \cite{Riess:2016jrr,Bernal:2016gxb,Huang:2018dbn}. Since neutrino masses require the introduction of at least two new fermionic degrees of freedom, this alternative constraint would not significantly loosen the bound quoted here.}
\begin{eqnarray}
\lambda_c < 10^{-9} \left(\frac{1}{\lambda} \right) \left(\frac{m_\phi}{1\,{\rm GeV}}\right)^2 \ .
\end{eqnarray}

It is important to keep in mind that $\lambda_c$ and $\lambda$ are qualitatively different couplings; $\lambda_c$ are marginal Yukawa couplings while $\lambda$ parameterize the consequences of higher-dimensional operators below the scale of electroweak symmetry breaking. Understanding whether it is reasonable to assume that $\lambda_c$ are tiny is akin to asking why the neutrino Yukawa couplings $y_{\nu}$ are $\mathcal{O}(10^{-12})$. Indeed, since $\lambda_c$ are also Yukawa couplings, the two types of interactions may be related in some mysterious way. Another concern is to ask whether we should have also included in the effective Lagrangian the dimension-six operators
\begin{equation}
\mathcal{L}_{\phi} \supset \frac{\nu^c_i \nu^c_j |H|^2 \phi^*}{(\Lambda^c)^2_{ij}} \ .
\label{eq:dim5c}
\end{equation}
After electroweak symmetry breaking, these modify $\lambda_c^{ij}\to \lambda^{ij}_c+v^2/(\Lambda^c)^2_{ij}$. The constraint above requires $\Lambda^c\gg \Lambda$ if one is interested in $\lambda$ values of order one (see Eq.~\eqref{EFToperators5}). Whether or not this is plausible depends on the ultraviolet physics that leads to the effective operators in question. We return to concrete examples in Sec.~\ref{sec:UV}.

%%%%%%%%%%

\section{Impact on neutrino-beam experiments}
\label{sec:oscillation}

Since $\phi$ couples predominantly to neutrinos via the coupling $\lambda$, its existence would impact neutrino scattering experiments. Furthermore, since it carries lepton number, $\phi$-emission leads to apparent lepton-number-violating effects in neutrino scattering. We discuss in some detail the physics of $\phi$-emission in neutrino--matter scattering, followed by current constraints and the sensitivity of next-generation neutrino scattering experiments, especially the LBNF-DUNE proposal. 

We will be most interested in $\phi$ from ``neutrino beamstrahlung'' and accelerator-based neutrino beams. We assume that constraints from atmospheric neutrinos are not competitive given the existing uncertainties on the atmospheric neutrino flux and the fact that the incoming neutrino direction is, a priori, unknown. We also assume that even in long-baseline, Earth-bound beam experiments, neutrino decays induced by $\phi$-exchange are negligible. This is a safe assumption; for $m_{\phi}\gg 1$~eV, we can estimate the lifetime for $\nu\to\bar{\nu}\nu\bar{\nu}$ (mass-eigenstate indices implicit) as 
\begin{equation}
\tau_{\nu} \sim \tau_{\mu}\left(\frac{m_{\mu}}{m_{\nu}}\right)^5\left(\frac{m_{\phi}}{v\lambda}\right)^4 \sim 2\times 10^{13}\left(\frac{1{\rm eV}}{m_{\nu}}\right)^5\left(\frac{m_{\phi}}{100{\rm MeV}}\right)^4\left(\frac{1}{\lambda}\right)^4~{\rm years},
\end{equation}
where $\tau_{\mu}, \, m_{\mu}$ are the muon lifetime and mass, respectively. Oscillation experiments with Earth-born neutrinos are sensitive to lifetime values shorter than nanoseconds~\cite{Coloma:2017zpg}. Even solar neutrino experiments are only sensitive to lifetimes shorter than a tenth of a millisecond~\cite{Beacom:2002cb,Berryman:2014qha}. The right-handed neutrino couplings $\lambda_c$ also mediate neutrino decay. Since we already assume these to be very small -- see Sec.~\ref{sec:lc} -- we ignore all their effects henceforth.

\subsection{General Discussion}

\begin{figure}[h]
\centerline{\includegraphics[width=0.4\textwidth]{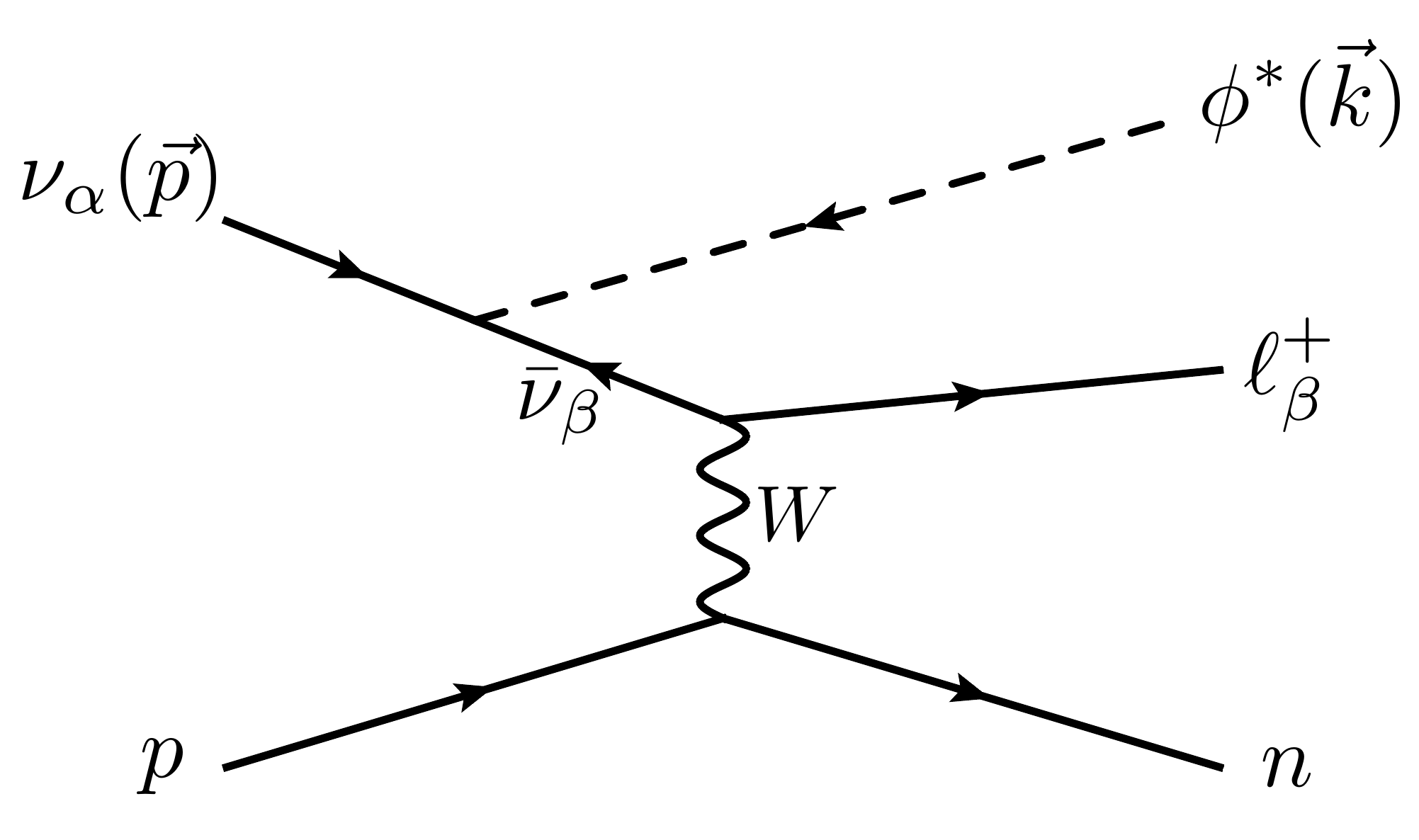} }
\caption{The Feynman diagram of interest for $\phi^*$ emission in neutrino scattering. The time direction is from left to right. A neutrino with flavor $\alpha$ emits a $\phi^*$, converting into an antineutrino with flavor $\beta$ before scattering off a nucleon and creating a positively charged lepton $l_\beta^+$. In the models we consider, $\phi^*$ decays invisibly into neutrinos.
}\label{PhiBrem}
\end{figure}

The Feynman diagram associated to a neutrino interacting with a nucleon target accompanied by $\phi$ radiation is depicted in Fig.~\ref{PhiBrem}. The amplitude for $\nu_{\alpha} + p \to \ell_{\beta}^+  +n+ \phi^*$ takes the general form
\begin{equation}\label{phi-process}
\begin{split}
\mathcal{A} &= \mathcal{A}_{CC} \frac{i}{\slashed{p}-\slashed{k}-m_\nu} (i\lambda_{\alpha\beta}) u_\nu(p) \\
&\simeq \lambda_{\alpha\beta} \mathcal{A}_{CC}\slashed{k}u_\nu(p) \frac{1}{2 p\cdot k - m_\phi^2} + \mathcal{O}(m_\nu) \ ,
\end{split}
\end{equation}
where $p$ is the four-momentum of the initial-state neutrino $\nu_\alpha$, $k$ is the four-momentum of the outgoing $\phi^*$, and $\mathcal{A}_{CC}$ is the amplitude for the antineutrino charged-current interaction $p\bar\nu_\beta\to n \ell^+_\beta$ with the antineutrino leg amputated. Superficially, the most striking signal here is the production of a wrong-sign charged lepton which could be identified in magnetized detectors, capable of distinguishing the electric charge of the final-state charged lepton.

\begin{figure}[h]
\centerline{\includegraphics[width=0.55\textwidth]{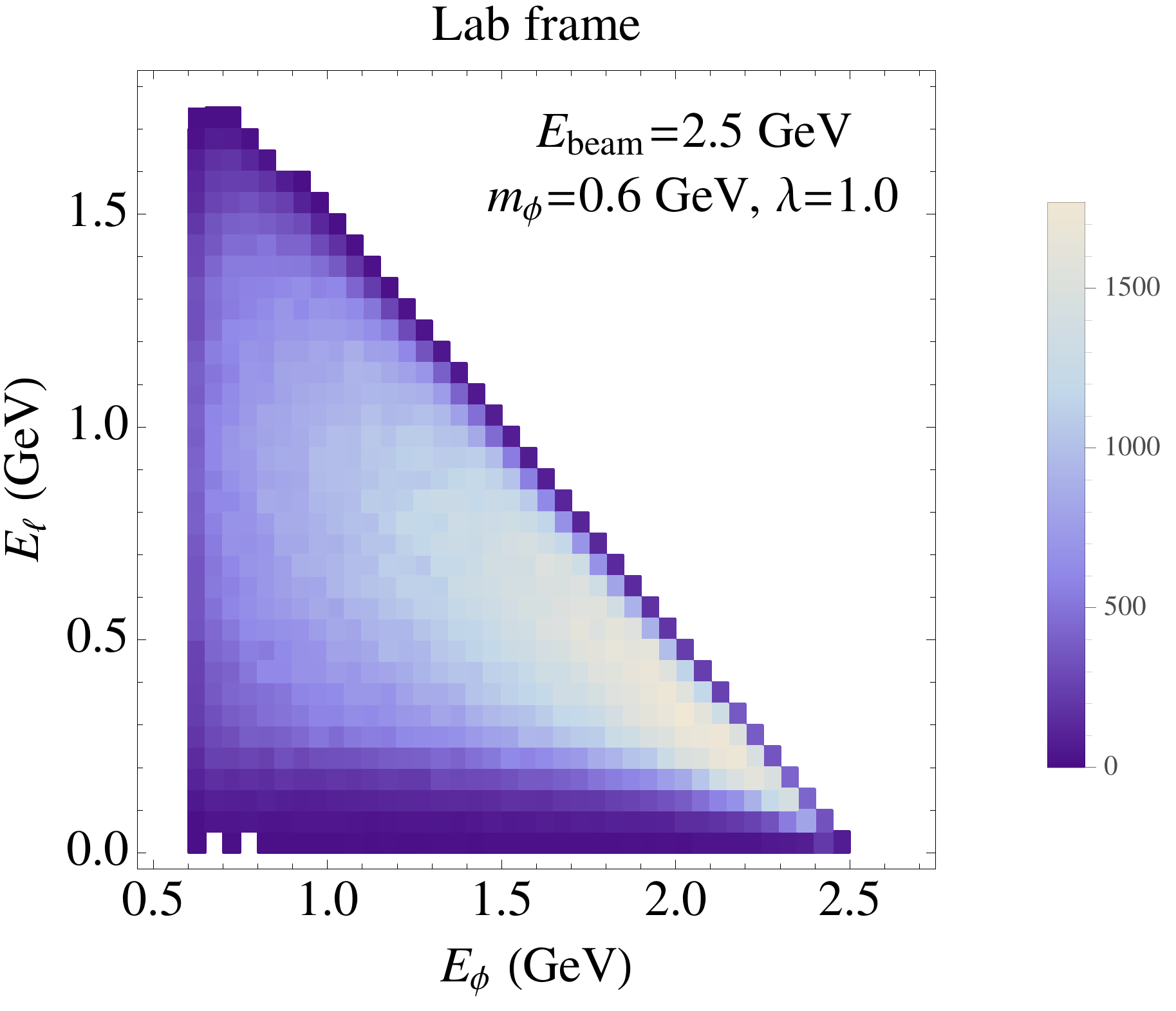} }
\caption{Correlated final state energy distribution of the charged lepton $\ell_\beta^+$ and the {\LeNCS} $\phi^*$, for a 2.5\,GeV neutrino beam striking a proton at rest, for $m_{\phi}=600$~MeV. The colorful legend bar shows the number of events falling in each $50\,$MeV$\times \, 50\,$MeV bin, out of half a million simulated events.}\label{fig:Dalitz}
\end{figure}

On the kinematics side, when $m_\phi \to 0$, the amplitude above is singular in the limit where the relative angle $\theta$ between $\vec{p}$ and $\vec{k}$ is zero. This corresponds to a collinear divergence of $\phi$ radiation. There is no infrared divergence in the limit $|\vec{k}|\to 0$. When $m_\phi$ is comparable to the neutrino beam energy, it is possible for $\phi$ to be radiated at large angles. As we will see, this can lead to sizable missing transverse momentum in a charged-current event and may provide a useful handle to identify the new physics contribution relative to SM backgrounds (see, e.g., Fig.~\ref{fig:DUNEEvts} for more details). 

Fig.~\ref{fig:Dalitz} depicts the final state energy distribution of the charged lepton $\ell_\beta^+$ and the {\LeNCS} $\phi^*$ for 2.5\,GeV neutrinos striking a proton target, assuming $m_{\phi}=600$~MeV. Clearly, for most of the events, the $\phi^*$ particle takes away most of the beam energy. Nonetheless, the charged lepton also typically carries enough energy to be measured in a neutrino detector.

Throughout, we use {\tt MadGraph5}~\cite{Alwall:2011uj} (with model file implemented using {\tt FeynRules}~\cite{Christensen:2008py}) to simulate the new physics as well as the SM processes, with the characteristic incoming neutrino beam energy spectrum for each experiment.  We also calculated the differential cross section for $\nu_{\alpha} + p \to \ell_{\beta}^+  +n+ \phi^*$ analytically (based on the $2\to3$ phase space integral given in~\cite{formcalc}) and found good agreement with the Monte Carlo simulations. In our calculations, for the neutrino energies of interest, we only consider the nucleon recoil, treating them as elementary particles. We find that this approximation is reasonable given our aspirations and will comment on it further in Sec.~\ref{subsec:miniboone}.

\subsection{MINOS}

The MINOS detector is magnetized, allowing charge identification between $\mu^+$ and $\mu^-$, and is hence sensitive to apparent changes in lepton number. The MINOS neutrino beam consists of 91.7\% $\nu_\mu$ and 7\% $\overline{\nu}_\mu$~\cite{Sousa:2015bxa, Rebel:2013vc}. With it, the collaboration measured the charged-current interaction rate for muon antineutrinos, $\overline{\nu}_\mu + p \rightarrow \mu^+ + n$, to be $3.84 \pm 0.05$ events/$10^{15}$ protons-on-target (POT)~\cite{Adamson:2012hp}.  A nonzero $\lambda_{\mu\mu}$ coupling leads to additional events with a $\mu^+$ in the final state associated to the muon neutrino flux --  roughly thirteen times greater than the antineutrino flux --  by radiating a $\phi$ particle, as depicted in Fig.~\ref{PhiBrem}. Defining the ratio of cross sections
\begin{equation}\label{eq:MINOSFrac}
\mathcal{R} = \frac{\sigma(\nu_\mu + p \rightarrow \mu^+ + \phi + n)}{\sigma(\overline{\nu}_\mu + p \rightarrow \mu^+ + n)} \ ,
\end{equation} 
and requiring that the additional contribution from Majoron emission does not modify the observed rate by more than $2\sigma$, we arrive at $\mathcal{R} \lesssim 0.002$. This implies that $|\lambda_{\mu\mu}| \lesssim 1$ for MeV $< m_\phi <$ GeV. Fig.~\ref{fig:NuResults}~(left) depicts the upper bound on $|\lambda_{\mu\mu}|$ as a function of $m_\phi$ -- black curve -- from MINOS. For simplicity, the cross sections in Eq.~\eqref{eq:MINOSFrac} are calculated for a constant neutrino (or antineutrino) energy of $3$ GeV, roughly in agreement with the peak of the experiment's neutrino spectrum.

\subsection{MiniBooNE}
\label{subsec:miniboone}

\begin{figure}
\centerline{\includegraphics[width=0.8\textwidth]{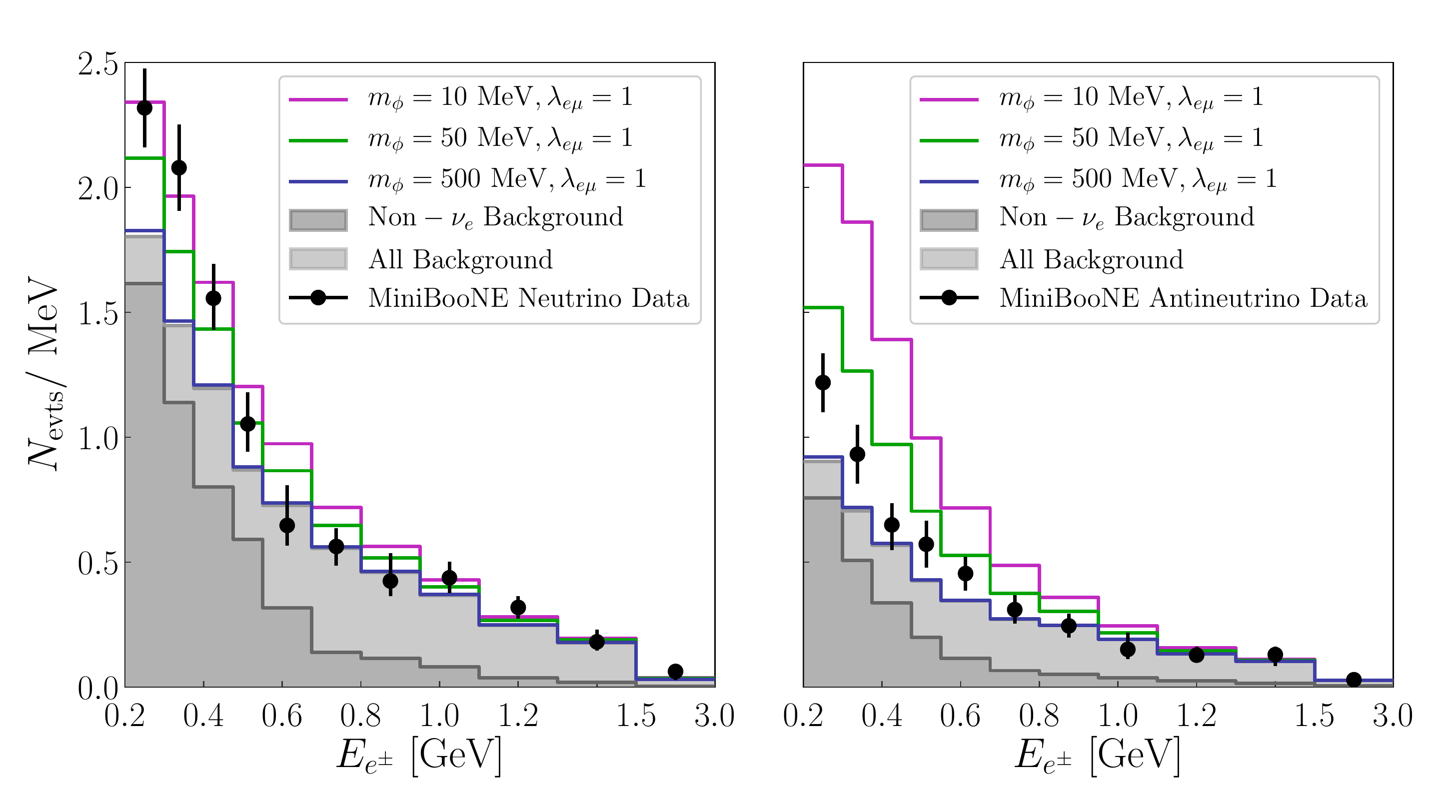}}
\caption{Simulated event yields at MiniBooNE as a function of the reconstructed electron/positron energy $E_{e^\pm}$. The left plot corresponds to running the experiment with the neutrino beam and the right plot to the antineutrino beam. We calculate the contributions to the signal of $\nu_\mu + p \to e^+ + n + \phi$ or $\bar \nu_\mu + n \to e^- + p + \phi^*$. The purple, green, and blue curves correspond to $m_\phi$ equal to 10, 50, 500\,MeV. All the curves assume $\lambda_{\mu e} = 1$, and the number of signal events simply scales as $|\lambda_{\mu e}|^2$. For comparison, the published data points are included with statistical error bars. The backgrounds are taken from~\cite{AguilarArevalo:2010zc, Aguilar-Arevalo:2013dva}, and the signals are stacked on top of the background. For simplicity, we assume the proton and neutron are elementary particles and ignore nuclear effects. Corroborating this approximation, we are able to reproduce the reported $e^+/e^-$ energy spectra -- light grey background histograms -- using the published beam $\nu_e$  and $\bar \nu_e$ energy spectra.
}
\label{fig:MB}
\end{figure}

The MiniBooNE beam consists mostly of muon-neutrinos plus a small electron-neutrino contamination~\cite{AguilarArevalo:2010zc, Aguilar-Arevalo:2013dva}. The experiment searched for electron-type events as a function of energy and reported a significant excess at energies below 400 MeV. This signal may be interpreted as a nonzero probability $P(\nu_\mu \rightarrow \nu_e)$ that a muon-type neutrino will be measured as an electron-type neutrino at short baselines, not inconsistent with data from the LSND experiment. A nonzero $\lambda_{\mu e}$, however, provides an additional source of electron-type events from the process $\nu_\mu + p \rightarrow e^+ + \phi^* + n$. Note that the overall rate is enhanced by the large ratio of $\nu_\mu$ to $\nu_e$ fluxes. Since MiniBooNE has no charge identification capabilities, such a signal is indistinguishable from $e^-$ appearance.

We simulate the expected spectrum of events as a function of the measured electron/positron energy for three values of $m_\phi$ -- $10$ MeV, $50$ MeV, and $500$ MeV -- and compare with the results from the MiniBooNE experiment. The simulation uses the reported MiniBooNE muon-neutrino flux as a function of neutrino energy. We also simulate the expected background using the MiniBooNE electron-neutrino flux to validate our approximation of treating the nucleons as elementary particles. These results are depicted in Fig.~\ref{fig:MB}. In order to explain the excess of events, we require a new-physics signal that is comparable in size to the $\nu_e$-beam induced background. Fig.~\ref{fig:MB} indicates that this can be achieved for $m_\phi \simeq 50$ MeV and $|\lambda_{\mu e}| \simeq 1$. While this serves as an attractive potential solution to the MiniBooNE low-energy excess, we note that this preferred region of parameter space is safely ruled out by the meson decay bounds discussed in Sec.~\ref{sec:meson} (see the yellow stars in Fig.~\ref{allconstraints}).

Similarly, one could try to explain the excess of $\bar{\nu}_e$ events reported by the LSND experiment assuming they are related to the new physics process $\nu_{\mu}+p\to n+e^++\bar{\nu}_e+\phi$. At LSND, however, the neutrino beam energies are of order tens of MeV and $\phi$-emission is only possible if $m_{\phi}\lesssim 10$~MeV. For such light masses, bounds on $\lambda$ from pion and kaon decays are too severe (Fig.~\ref{allconstraints}) for these effects to resolve the LSND anomaly. Furthermore, the $\nu_{\mu}$ beam at LSND is quasi-monochromatic ($E_{\nu}\sim 30$~MeV), so the intermediate antineutrinos that mediate $\nu_{\mu}+p\to n+e^++\phi^*$ have energies below 30 MeV. The LSND excess extends to reconstructed neutrino energies that exceed this bound. 

\subsection{NOMAD}

Between 1995 and 1998, the NOMAD experiment at CERN searched for oscillations of $\nu_\mu$ (and $\nu_e$) into $\nu_\tau$~\cite{Astier:2001yj}. The neutrino energies and baseline were such that NOMAD was sensitive to small mixing angles but mass-squared differences that are much larger than those that were ultimately revealed by neutrino oscillation experiments. NOMAD did not observe any $\nu_\tau$ CC events and placed an upper limit on the oscillation probability $P(\nu_\mu \rightarrow \nu_\tau) < 2.2 \times 10^{-4}$. Using the published NOMAD neutrino flux and simulated scattering cross section, we can translate this limit into one on the coupling $|\lambda_{\mu\tau}|$ as a function of $m_\phi$ (similar to the procedure discussed above with MINOS). Since the NOMAD beam consisted of high-energy neutrinos, this limit extends to large values of $m_\phi\gg 1$~GeV, despite the necessary energy budget required to produce both a $\phi$ and a $\tau^+$. The resulting limit is depicted in Fig.~\ref{fig:NuResults}~(right), black curve. Similar bounds can also be extracted from the CHORUS experiment. Their results~\cite{Eskut:2007rn}, however, as far $P(\nu_\mu \rightarrow \nu_\tau)$ is concerned, are slightly less sensitive than those from NOMAD.

A bound on $|\lambda_{e\tau}|$ can be similarly extracted from the NOMAD bound $P(\nu_e \rightarrow \nu_\tau)\lesssim10^{-2}$ \cite{Astier:2001yj}. We expect the extracted bound to be at least one order of magnitude weaker than the one on $|\lambda_{\mu\tau}|$ discussed above and hence outside the region of the parameter space depicted in Fig.~\ref{fig:NuResults}~(left).  

\subsection{DUNE}

The upcoming Deep Underground Neutrino Experiment (DUNE) will consist of both a near detector and a far detector. The former is expected to collect $\mathcal{O}(10^5)$ $\nu_\mu$ CC events per year~\cite{Acciarri:2015uup}. Similar to the MINOS discussion, we could use the channel $\nu_\mu + p \rightarrow \mu^+ + \phi^* + n$ to search for nonzero $\lambda_{\mu\mu}$.  At the DUNE near detector, however, it is anticipated that the charge of the muon will not be identified and hence one needs to account for several wrong-sign background processes, including $\nu_\mu + n \rightarrow \mu^- + p$. 

\begin{figure}
\centerline{\includegraphics[width=0.45\textwidth]{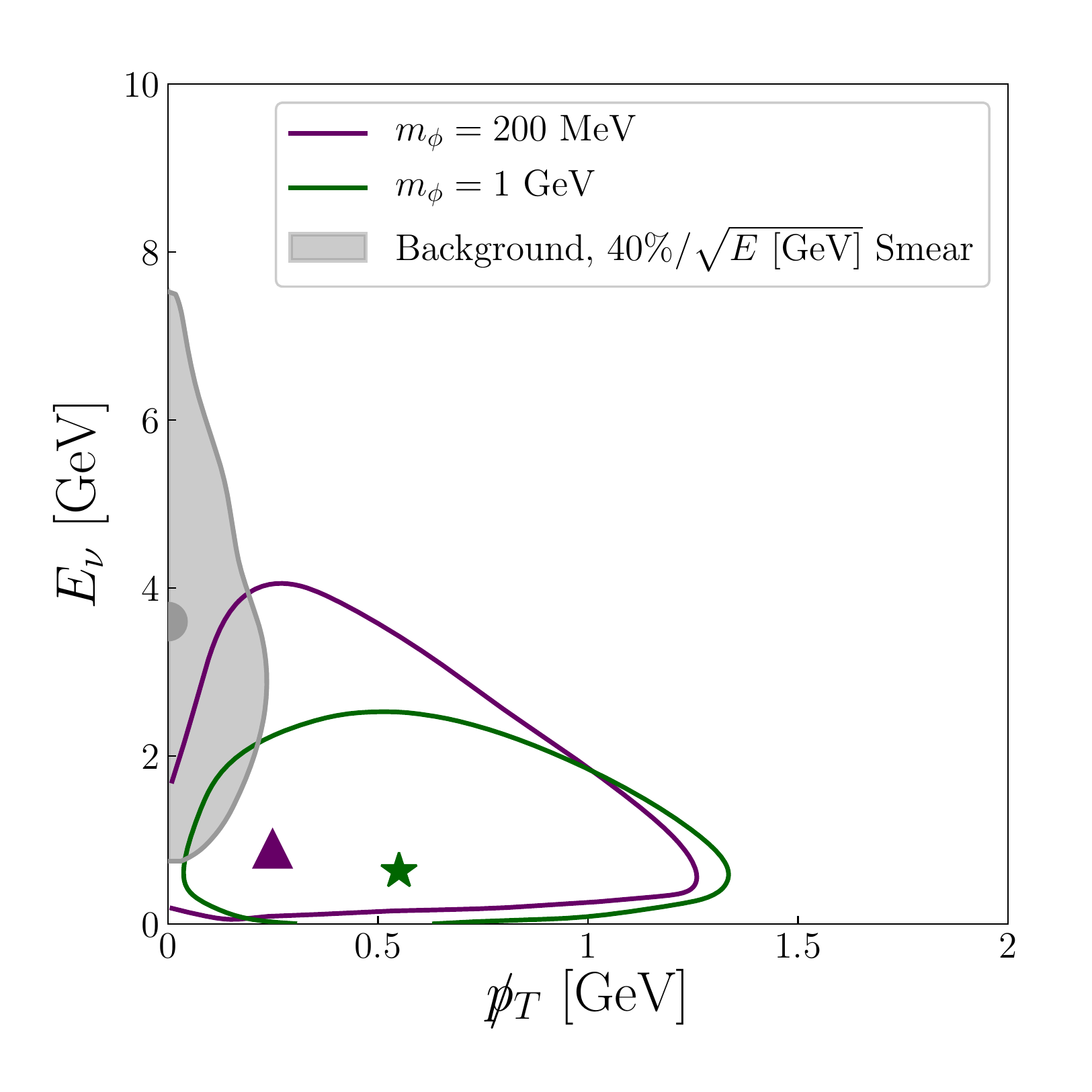}\includegraphics[width=0.45\textwidth]{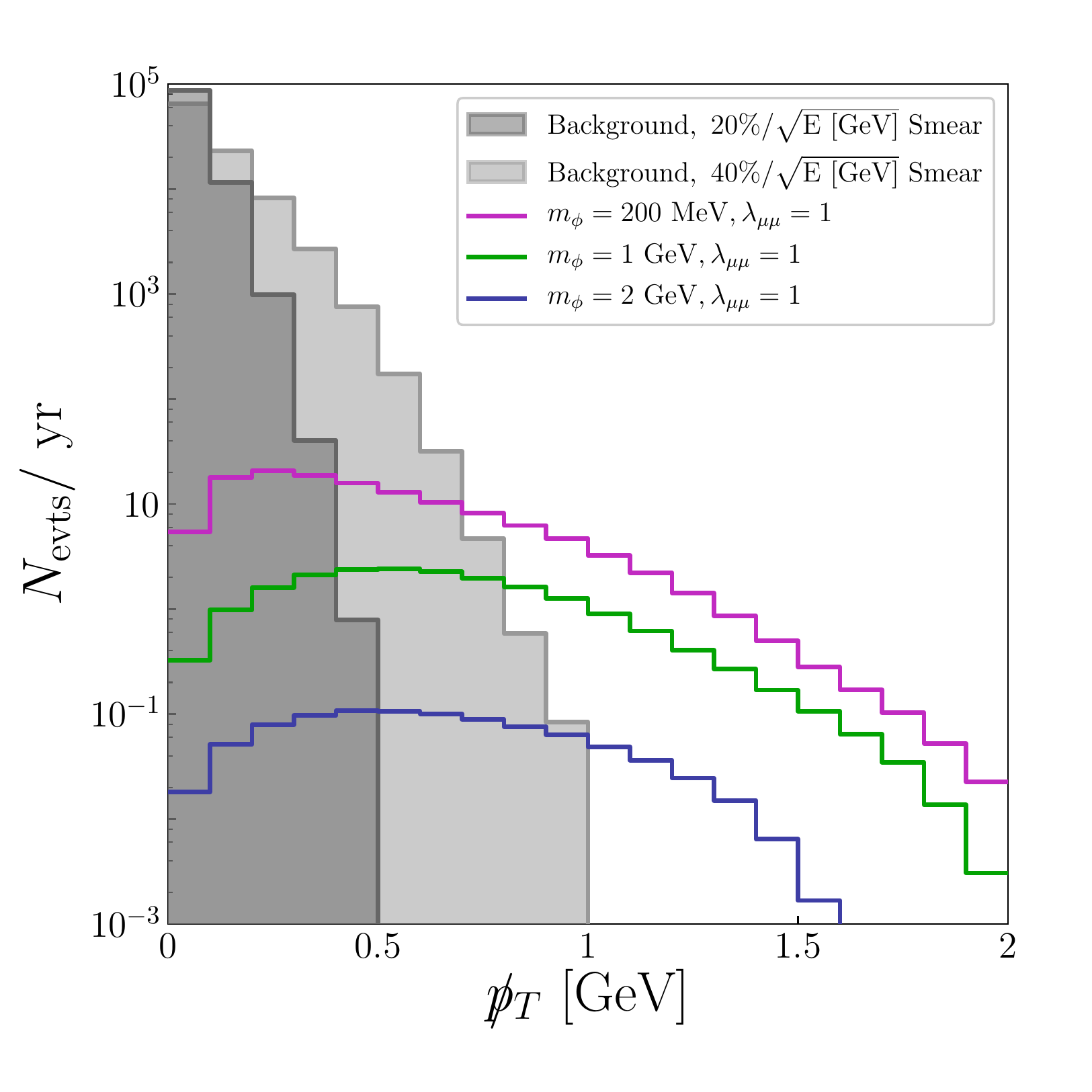}}
\caption{{\sf Left}: Region of the $\slashed{p}_T\times E_\nu$-plane containing most of the simulated signal and background events. The solid lines encompass 90\% of the events for the signal with $m_\phi = 200$ MeV (purple) and $1$ GeV (green), and for the SM background $\nu_\mu + n \rightarrow \mu^- + p$ (black). Corresponding markers for each of these show the peak of the event distribution. {\sf Right}: The number of events per year in a given bin of $\slashed{p}_T$, after marginalizing over $E_\nu$, for the background (dark: $20\%/\sqrt{E\ \mathrm{[GeV]}}$ hadronic energy resolution, light: $40\%/\sqrt{E\ \mathrm{[GeV]}}$ hadronic energy resolution) and the signal (purple: $m_\phi = 200$ MeV, green: $m_\phi = 1$ GeV, blue $m_\phi = 2$ GeV, all with $\lambda_{\mu\mu} = 1$) at the DUNE near detector, assuming $10^5$ background events per year. We see that, for the more optimistic hadronic energy resolution, there are no background events with $\slashed{p}_T > 0.5$ GeV and for the less optimistic case, there are no background events with $\slashed{p}_T > 1$ GeV.}
\label{fig:DUNEEvts}
\end{figure}

In order to address this issue, we explore the kinematics of the final state, especially the presence of missing transverse momentum  $\slashed{p}_T$ when $\phi$ is radiated at large angles. The SM background $\nu_\mu + n \rightarrow \mu^- + p$ has no $\slashed{p}_T$ assuming all final-state particles can be reconstructed with good precision. To estimate this effect, we simulate the background channel assuming energy reconstruction uncertainties between $20-40\%/\sqrt{E\ \mathrm{[GeV]}}$ for the outgoing proton and $3\%/\sqrt{E\ \mathrm{[GeV]}}$ for the outgoing muon.\footnote{The same background channel associated with the antineutrino beam ($\bar{\nu}_\mu + p \rightarrow \mu^+ + n$) is expected to have much larger energy reconstruction uncertainties because of the neutron in the final state. For this reason we concentrate on the neutrino-beam configuration as opposed to the antineutrino one.} In the left panel of Fig.~\ref{fig:DUNEEvts}, we display the region of the $\slashed{p}_T\times E_\nu$-plane\footnote{The inferred neutrino energy assumes a $2\to 2$ scattering process of a neutrino off an at-rest nucleon. The inferred signal neutrino energy, given this incorrect assumption, does not match the real incoming neutrino energy.} that contains most of the signal and background events. The solid lines encompass 90\% of the events for the signal with $m_\phi = 200$ MeV (purple) and $1$ GeV (green), and for the SM background $\nu_\mu + n \rightarrow \mu^- + p$ (black). Corresponding markers for each of these show the peak of the event distribution. As expected, signal events tend to have higher $\slashed{p}_T$ and lower $E_\nu$, particularly for large values of $m_\phi$. In this two-dimensional space, one can define a cut to optimize the sensitivity to the signal. For simplicity, however, we will consider the projection of this distribution down to the $\slashed{p}_T$-axis, as depicted in Fig.~\ref{fig:DUNEEvts}~(right). Here, we assume $10^5$ background events per year.

\begin{figure*}
\centerline{\includegraphics[width=0.9\textwidth]{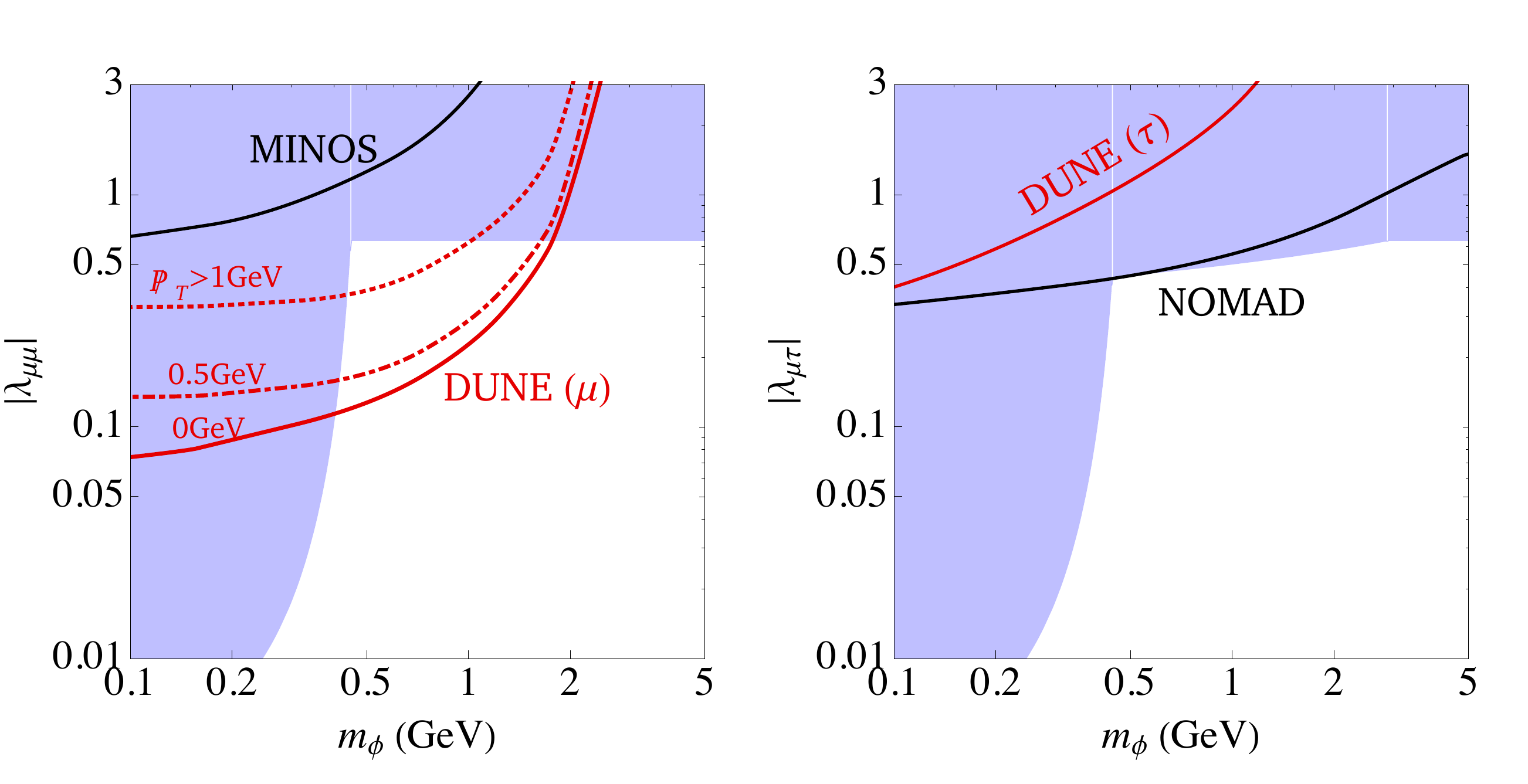}}
\caption{Sensitivity of existing and future oscillation experiments to $\phi$ beamstrahlung in the $\lambda \times m_\phi$-plane.
The light blue shaded region is the union of all the constraints summarized in Fig.~\ref{allconstraints}. 
Existing lower limits from MINOS and NOMAD are shown by the black curves. 
The future DUNE experiment could further probe $\lambda$ values as low as indicated by the red curves. 
The red solid, dot-dashed, and dotted curves corresponds to a missing transverse momentum cut $\slashed{p}_T$ greater than $0, 0.5, 1$\,GeV, respectively, 
for the signal event selection. See text for details. 
}
\label{fig:NuResults}
\end{figure*}

We perform three different analyses: one assuming zero background (to establish the most optimistic result), one assuming $20\%/\sqrt{E\ \mathrm{[GeV]}}$ resolution for the final-state protons, for which we impose a $\slashed{p}_T > 0.5$~GeV cut, and one assuming $40\%/\sqrt{E\ \mathrm{[GeV]}}$ energy resolution, for which we impose a $\slashed{p}_T > 1$~GeV cut. We calculate the value of $|\lambda_{\mu\mu}|$ for which DUNE can detect, above the given $\slashed{p}_T$ cut, one signal event per year, or 10 signal events over the experimental run of 10 years. Assuming no background, we postulate this amounts to a discovery. The result is depicted in Fig.~\ref{fig:NuResults}~(left), red lines. The most optimistic line is solid, whereas the dot-dashed line corresponds to the $\slashed{p}_T > 0.5$ GeV cut, and the dashed line corresponds to the $\slashed{p}_T > 1$ GeV cut. The sensitivity of the DUNE experiment surpasses the currently disfavored region -- light blue area -- for $m_\phi$ values between roughly $300$~MeV and $2$~GeV.

When running in neutrino mode, DUNE still has a nonzero $\bar{\nu}_\mu$ component to its flux, on the order of 10\% of the total beam makeup. These $\bar{\nu}_\mu$ would contribute to the background, however the final state would include a (harder-to-reconstruct) neutron as well as a $\mu^+$. This irreducible background will impact the sensitivity depicted in Fig.~\ref{fig:NuResults}~(left). We expect that the number of events per year due to this background would be roughly a factor of $20$ lower than the light grey curve shown in the right panel of Fig.~\ref{fig:DUNEEvts} -- a factor of $10$ due to the 10\% of the beam, and a factor of two because the antineutrino-nucleon cross section is smaller than the neutrino-nucleon one. We expect, in the absence of a signal, that the experimental collaboration will be able to set a limit somewhere between the $\slashed{p}_T>0.5$~GeV and the $\slashed{p}_T>1$~GeV curves. A more thorough analysis using both $\slashed{p}_T$ and $E_\nu$, of course, could improve this sensitivity.

In the near detector, a nonzero $\lambda_{\mu\tau}$ would lead to wrong-sign $\tau$-appearance, $\nu_\mu + p \rightarrow \tau^+ + n + \phi^*$. Because the $\tau^+$ is difficult to identify in the detector and requires high-energy neutrinos in order to be produced, we expect the capability of DUNE to detect this to be significantly worse. To determine the range of $\lambda_{\mu\tau}$ to which DUNE will be sensitive, we simulate data and estimate the values for which there would be one $\tau^+$ event in the near detector each year, whether or not that event is properly identified. The resulting curve, as a function of $m_\phi$, is depicted as a red line in Fig.~\ref{fig:NuResults}~(right). Even in a perfect world, DUNE will not be able to improve on current bounds from NOMAD, Z-boson decays, and Higgs decays.

DUNE will also be capable of producing $\phi$'s with nonzero $\lambda_{\mu\tau}$ and $\lambda_{\tau\tau}$ at the far detector because, given the $1300$~km baseline, the oscillation probability $P(\nu_\mu \rightarrow \nu_\tau)$ is large for the energies of interest. The $\nu_\tau$ can then interact via  $\nu_\tau + p \rightarrow \mu^+ + n + \phi^*$ (with $\lambda_{\mu\tau}$) or $\nu_\tau + p \rightarrow \tau^+ + n + \phi^*$ (with $\lambda_{\tau\tau}$). Both of these would result in large values of $\slashed{p}_T$, similar to the near-detector discussion above, particularly in the latter case if the $\tau^+$ subsequently decays to a $\mu^+$ and neutrinos. However, at the far detector, the dominant background for this is $\nu_\tau + n \rightarrow \tau^- + p$, where, if the $\tau^-$ decays to a $\mu^-$, this background will have a signature similar to the signal in terms of its missing transverse momentum distribution. We find that this background completely dominates the proposed signal for values of $m_\phi$ and $\lambda_{\mu\tau}$ or $\lambda_{\tau\tau}$ of interest.

\bigskip 
The currently-running NuMI Off-Axis $\nu_e$ Appearance (NO$\nu$A) experiment, as far as this analysis is concerned, is similar to DUNE, particularly in its near detector. The neutrino source for NO$\nu$A generates neutrinos with energies near $2$ GeV, and the experiment has a near detector that collects $\mathcal{O}(10^4-10^5)$ events per year. It can also search for high-$\slashed{p}_T$ events in its near detector, as proposed here. We expect the corresponding line in Fig.~\ref{fig:NuResults} for NO$\nu$A  to be less competitive due to the lower event rate and slightly worse energy reconstruction resolution compared to those expected for DUNE. We do not attempt a detailed estimate of the sensitivity here but strongly encourage searches for high $\slashed{p}_T$ events at the NO$\nu$A near detector.

\section{Possible Ultraviolet Completions}
\label{sec:UV}

In this section, we discuss possible ultraviolet-complete models that, after integrating out heavy degrees of freedom, lead to the dimension-six operator in Eq.~\eqref{EFToperators5}. All models discussed are inspired by the tree-level realizations of the seesaw mechanism; the main difference here is that all new particle properties as well as their interactions are such that the {\BmL} symmetry is preserved.

One option is to introduce a scalar $T$, a triplet under $SU(2)_L$ with hypercharge $+1$ and {\BmL} charge $+2$. We will call it the type II model, because it has a structure similar to the type-II seesaw. As already highlighted, however, unlike the seesaw mechanism, there are no {\BmL}-violating effects here. The most general renormalizable Lagrangian in this case contains
\begin{equation}
\label{TInt}
\mathcal{L}_{\rm UV} \supset \tilde y_{\alpha\beta} L_\alpha T L_\beta + \lambda_{T} HT^\dagger H  \phi - M_T^2 {\rm Tr}(T^\dagger T)  + {\rm h.c.} \ ,
\end{equation}
where $\tilde y_{\alpha\beta}$ are Yukawa couplings between the triplet $T$ and leptons of flavor $\alpha$ and $\beta$, $\lambda_{T}$ are scalar couplings between the triplet, the Higgs field and the {\LeNCS} $\phi$, and $M_T$ is the triplet scalar mass. When the $T$ field is integrated out, the low-energy effective theory matches that in Eq.~\eqref{EFToperators5} with
\begin{equation}
\frac{1}{\Lambda_{\alpha\beta}^2} = \frac{\tilde y_{\alpha\beta}\lambda_T}{M_T^2} \ .
\end{equation}
In this scenario, the effective operator allowing interactions of right-handed neutrinos with the {\LeNCS}, discussed in Eq.~\eqref{eq:dim5c}, is not generated at the tree-level and one expects $\Lambda\ll \Lambda^c$. It is, therefore, technically natural to have the {\LeNCS}  more strongly coupled to left-handed neutrinos than to right-handed ones, as discussed in Sec.~\ref{sec:lc}. 

Another option, which we call the type I model, is to introduce pairs of vector-like fermions $N_i$ and $N_i^c$ ($i=1,2,\ldots,n$, the number of vector-like fermions) that are SM singlets carrying {\BmL} charges $\mp1$, respectively. The most general renormalizable Lagrangian includes
\begin{equation}
\label{NNcInt}
\mathcal{L}_{\rm UV} \supset \tilde y_{\alpha i} L_{\alpha i} H N_i^c  +  M_{N,i} N_i N_i^c + \lambda_{N, ij} \phi N_i N_j +  \lambda^c_{N, ij} \phi^* N^c_i N^c_j +  \tilde{\lambda}^c_{N\nu, ij} \phi^* N^c_i \nu^c_j + {\rm h.c.} \ ,
\end{equation}
where $\tilde y$ are the strengths of the new Yukawa interactions and $\lambda_{N}$ characterizes the strength of the interaction between $N^c$ and the {\LeNCS} field $\phi$.\footnote{We omit terms proportional to $N_i\nu_j^c + {\rm h.c.}$. This can be done without loss of generality and consists of a ``basis-choice'' for the $N^c,\nu^c$ fields.} The constraint that the right-handed neutrino couplings $\lambda^{ij}_c$ to $\phi$ are very small -- see Sec.~\ref{sec:lc} -- implies that  $\lambda^c_{N, ij}$ and $\tilde{\lambda}^c_{N\nu, ij}$ are also small and henceforth neglected. When all heavy fermion fields are integrated out, we obtain the effective operator in Eq.~\eqref{EFToperators5}, $(L_\alpha H)(L_\beta H)\phi/\Lambda_{\alpha\beta}^2$, with 
\begin{equation}\label{eq:UVrelation}
\frac{1}{\Lambda_{\alpha\beta}^2} = \sum_{i,j} \tilde y_{\alpha i} \frac{1}{M_{N_i}} \lambda_{N, ij} \frac{1}{M_{N_j}} \tilde y_{\beta j} \ .
\end{equation}
Here, after electroweak symmetry breaking, the SM neutrinos mix with the new vector-like fermions. The mass matrix in the $\{\nu, \, N\}\times\{\nu^c, \, N^c\}$ basis takes the form 
\begin{equation}
\begin{pmatrix}
\nu & N
\end{pmatrix}
\begin{pmatrix}
y_\nu v/\sqrt2 &  \tilde y v/\sqrt2 \\
0 & M_N
\end{pmatrix}
\begin{pmatrix}
\nu^c \\ 
N^c
\end{pmatrix} \ .
\end{equation}
After diagonalization, the active--sterile mixing angles between the $\nu$ and the $N$ fields are of order $\theta_{as}\sim \tilde{y}v/M_N$ and if $\Lambda$ is to be of order the weak scale, even for large $\lambda_{N, ij}$, we expect $\theta_{as}$ values to be of order one, clearly ruled out (see, for example, \cite{Atre:2009rg,Alonso:2012ji,Drewes:2015iva,deGouvea:2015euy,Fernandez-Martinez:2015hxa}). It may, however, be possible to finely tune the model by taking advantage of its flavor structure, a possibility we do not explore further here. It also behooves us to highlight that while large $\tilde{y}$ Yukawa couplings are required, the neutrino Yukawa couplings $y_{\nu}$, which are very similar as far as the symmetry structure of the theory is concerned, are much smaller, of order $10^{-12}$. 

Alternatively, the vector-like fermions $N_{ia}$ and $N_{ia}^c$ introduced above can be replaced by $SU(2)_L$ triplets, where $a$ is the $SU(2)_L$ index in the adjoint representation -- the type III model. Their Yukawa couplings take the form $\tilde y_\alpha L_{\alpha i} \sigma^a H N_{ia}$, where $\sigma^a$ are the Pauli matrices. In this case, when the $N_{ia}$ and $N_{ia}^c$ fields are integrated out, the low energy effective operator has the form $(L_\alpha \sigma^a H)(L_\beta \sigma^a H)\phi/\Lambda_{\alpha\beta}^2$ and $1/\Lambda_{\alpha\beta}^2$ is related to the ultraviolet parameters in the same way as Eq.~\eqref{eq:UVrelation}. In this scenario, there are no equivalent $\tilde{\lambda}^c_{N\nu, ij}$ couplings but the concerns raised above remain. 

All in all, both the type-I and type-III ultraviolet completions appear to be unsuccessful, while the type-II model works very well.  

As discussed earlier, we have been interested, for the most part, in effective couplings $\lambda_{\alpha\beta}=v^2/\Lambda^2_{\alpha\beta}$ (defined in Eq.~\eqref{Lint}) of order $\mathcal{O}(0.1-1)$. This implies that the mass scales of the new particles $T$ (or $N, \, N^c$) should be around the electroweak scale, unless one resorts to large (potentially non-perturbative) $\tilde y$ and $\lambda_{T,N}$ couplings. The Large Hadron Collider and future collider experiments have, therefore, the opportunity to probe specific ultraviolet completions of the scenario discussed here~\cite{Melfo:2011nx, Perez:2008ha}. On the other hand, as long as none of the new particles are lighter than the Higgs boson or the $Z$-boson, our discussions in Sections II and III, based on the effective field theory described by Eq.~\eqref{EFToperators5}, remain valid.

%%%%%%%%%%%%%%%%%%%%%%%%%%%%%%%%%%%%%%%%%%
%%%%%%%%%%                       New Section V                  %%%%%%%%%%%%
%%%%%%%%%%%%%%%%%%%%%%%%%%%%%%%%%%%%%%%%%%

\section{Dark Matter Connection}
\label{sec:DM}

As discussed in the introduction, all the gauge-invariant, Lorentz-invariant effective operators constructed out of the SM particle content plus any number of right-handed neutrinos carry even {\BmL}. This implies that if one extends the scalar sector of the SM by introducing {\LeNCS} fields (scalar fields with nonzero {\BmL} charge), the Lagrangian of the SM plus the {\LeNCS} fields could respect some accidental discrete symmetry; the {\LeNCS} could be stable and hence be an interesting dark matter candidate.

If one were to extend the SM with any number of {\LeNCS} species with integer {\BmL} charges $q_{B-L}$, then the Lagrangian would be invariant under a $Z_2$-symmetry where all SM fields are invariant and each {\LeNCS} species has $Z_2$-charge $(-1)^{q_{B-L}}$. In this scenario, the lightest odd-charged {\LeNCS} is stable.\footnote{Amusingly, it is easy to check that this $Z_2$ charge is a non-supersymmetry version of the $R$-parity charge, $(-1)^{3q_{B-L}+2s}$, where $s$ is the spin quantum number.} Concretely, if there is only one {\LeNCS} field $\chi$ with {\BmL} charge $+1$, then all of its couplings to SM fields involve operators that contain $\chi^{m}(\chi^*)^{2n-m}$ (plus their hermitian conjugates), where $n$, $m$ are integers ($m\le 2n$).  All of these operators are invariant under $\chi\leftrightarrow -\chi$ and this $Z_2$-symmetry implies that $\chi$ is stable and a simple, elegant dark matter candidate. The model here could also be viewed as an example of the longstanding wisdom that accidental symmetries can emerge once a model is endowed with higher symmetries~\cite{Georgi:1981pu, Aulakh:2000sn, Cheung:2015mea, Duerr:2017amf}. In this section, we will briefly discuss a few consequences of a {\LeNCS} dark matter candidate, including a dark sector interacting via a {\LeNCS} portal, and comment on some generic features. A detailed discussion of a subset of these possibilities was recently presented in Refs.~\cite{Bandyopadhyay:2017bgh,Cai:2018nob} (see also references therein).

Our discussions here are based on a simple model. In addition to the scalar $\phi$ with {\BmL} charge +2 discussed in the preceding sections, we introduce another SM singlet scalar field $\chi$ carrying {\BmL} charge $-1$, which serves as the dark matter candidate. The scalar potential of $\chi$, $\phi$ and the Higgs boson contains the following interaction-terms
\begin{equation}\label{DMint}
\mathcal{L} \supset \left(\mu_{\phi\chi} \phi \chi^2 + {\rm h.c.}\right) + c_{\phi\chi} |\phi|^2 |\chi|^2 + c_{H\chi} |H|^2 |\chi|^2 + \left(\chi^2 \hat{O}_{B-L=2} + {\rm h.c.} \right) + \cdots \ .
\end{equation}
$ \hat{O}_{B-L=2}$ are gauge-invariant operators constructed out of the SM particle content plus the right-handed neutrinos that carry {\BmL} charge equal to two. The ellipses represent higher-dimensional operators which couple $\chi^{2n}$ to SM operators carrying ${\text \BmL}=2n$, for $n>1$. The interactions in Eq.~\eqref{DMint} will determine the thermal relic abundance of $\chi$ and the relative importance of the different interactions govern how $\chi$ will manifest itself today. We have identified a few distinct scenarios. 
\begin{itemize}
\item {\it Neutrinophilic dark matter and a $\phi$-portal to the dark sector}. \ If the $\mu_{\phi\chi}$ and $c_{\phi\chi}$ couplings defined in Eq.~\eqref{DMint} mediate the most significant interactions controlling the physics of $\chi$ the $\phi$ particle plays the role of mediator between the SM sector and the dark matter $\chi$. 
Following the assumptions behind Eq.~\eqref{Lint}, where $\phi$ interacts with the SM sector mainly through the $(LH)^2 \phi$ dimension-six operator, $\chi$ is ``neutrinophilic.'' For example, two $\chi$ particles will annihilate into two SM neutrinos via tree-level $\phi$-exchange. {\BmL} conservation implies one cannot close a neutrino loop and use $Z$-boson exchange to couple $\chi$ to nucleons or charged leptons in the SM sector at the one-loop level. The lowest order contributions to such interactions occur at the two-loop level or higher. Under these circumstances, one expects a large hierarchy between the dark matter interaction strength to neutrinos relative to the other SM particles. This is different from other ``leptophilic'' dark matter candidates, proposed in Ref.~\cite{Fox:2011fx}. 
There are cosmological constraints on dark-matter--neutrino interactions from the CMB~\cite{Serra:2009uu,Wilkinson:2014ksa}. On the other hand, 
a sizable dark-matter--neutrino interaction cross section may prove useful for understanding small-scale structure formation in the universe~\cite{Bringmann:2013vra, Bertoni:2014mva}. 
\item {\it Higgs portal dark matter}. \ 
If, on the other hand, the $c_{H\chi}$ coupling defined in Eq.~\eqref{DMint} mediates the most significant interaction controlling the physics of $\chi$, $\chi$ is indistinguishable from a vanilla Higgs portal scalar dark matter candidate. In this case, imposing the correct thermal relic abundance for $\chi$ and satisfying the  latest direct-detection constraints leads to $\chi$ masses larger than a few hundred GeV~\cite{Escudero:2016gzx, Feng:2014vea, Cline:2013gha}. Measurements of the Higgs portal interaction are insensitive to -- and hence cannot reveal -- the {\BmL} charge of $\chi$. 
\item {\it Dark matter triggers nucleon decay}. \ If $\chi$ interacts with the SM mainly via higher-dimensional operators where the SM fields carries both $B$ and $L$ number -- contained in the last term in Eq.~\eqref{DMint} -- then it is possible for the dark matter to catalyze the decay of a nuclear neutron into a neutrino, $\chi + (Z,A) \to \chi^* + (Z,A-1) + \nu$. Even though the final state neutrino is invisible, the resulting hadronic activity due to the removal of a neutron from the target nucleus could lead to visible signatures if such a process occurs in a dark matter or a neutrino detector. For a recent phenomenological study, see Ref.~\cite{Huang:2013xfa}. 
\end{itemize}

\section{Concluding Remarks}

We explored the hypothesis that $B-L$ is a conserved global symmetry of nature and that there are new SM gauge-singlet scalar fields with integer nonzero $B-L$ charge. These lepton-number-charged scalars were dubbed {\LeNCS}, for short. Under this hypothesis, neutrinos are massive Dirac fermions and (at least two) right-handed neutrino fields also exist. We concentrated our discussion to a {\LeNCS} field $\phi$ with $B-L$ charge equal to two. At the renormalizable level, $\phi$ only couples to the SM Higgs boson and to the right-handed neutrinos. At the nonrenormalizable level, $\phi$ also couples to left-handed neutrinos via the Lagrangian spelled out in Eq.~\eqref{EFToperators5}, at the dimension-six level.

We find that, for masses below a GeV, a {\LeNCS} could first manifest itself at intense neutrino beam experiments via $\phi$-radiation: $\nu_{\alpha}+N\to \ell^++\phi^* + N'$, where $N, \, N'$ are different nuclei and $\alpha,\ell=e,\mu,\tau$. We compiled the relevant constraints from a variety of laboratory processes, along with future sensitivities, on the $\phi$--left-handed neutrino effective couplings $\lambda$ (see Eq.~(\ref{Lint})) in Figs.~\ref{allconstraints} and \ref{fig:NuResults}. While the $\phi$--right-handed neutrino couplings are only very poorly constrained in the lab, bounds on the number of relativistic species constrain it to be very small if the $\phi$--left-handed neutrino effective couplings $\lambda$ are accessible to next-generation neutrino beam experiments. Note that while  we situated our discussion in a scenario where $B-L$ conservation provides guidance concerning the structure of the new-physics Lagrangian, any scalar field that interacts predominantly via the operators contained in Eq.~\eqref{EFToperators5} will be subject to the same constraints. 

We concentrated on effective operators with mass-dimension less than or equal to six. At dimension eight, there are many more operators, including those involving quarks. All are listed in Appendix~\ref{app:dim8}. The effective scales of many dimension-eight operators are constrained to be very large for low-mass {\LeNCS} fields because these mediate the apparent violation of baryon number, including nucleon decays. For example, if $\phi$ is much lighter than a GeV, the operator number 10 in Table~\ref{table:dim8ops} in Appendix~\ref{app:dim8} -- the operator $ u^c d^c d^c (L H) \phi $ -- leads to a nucleon lifetime of order 
\begin{equation}
\tau_{n}\sim 10^{29}~{\rm years}\left(\frac{\Lambda_8}{10^8~\rm GeV}\right)^8~,
\end{equation}
where $\Lambda_8$ is the effective scale of the operator. For heavier $\phi$ masses, one expects less severe but still relevant bounds that may depend on other $\phi$ couplings. For example, if the $\lambda$ couplings discussed here are significant then operator number 10 in Table~\ref{table:dim8ops} leads to the neutron decay $n\to \nu \overline{\nu} \overline{\nu}$, mediated by off-shell $\phi$ decay. Whether it is consistent to have $\lambda$ of order one while $\Lambda_8$ is much larger than a TeV depends on the ultraviolet physics responsible for the effective theory. The type II example discussed in Sec.~\ref{sec:UV} only generates operators in Table~\ref{table:dim8ops} that mediate apparent lepton-number violation, not baryon-number violation. The reason is that, at the renormalizable level, there are no couplings between the heavy $T$ (colorless, $SU(2)$ triplets) and $SU(3)$ colored degrees of freedom. 

If $B-L$ is an exact symmetry of nature, odd-charged {\LeNCS} species are interesting dark matter candidates and only couple to SM degrees of freedom in pairs. The stability of odd-charged {\LeNCS} fields relies strongly on $B-L$ conservation. If, for example, {\BmL} is broken by quantum gravity effects, we would naively expect effective operators like  $\chi LH\nu^c/{M_{\rm P}}$, where $\chi$ has $B-L$ charge one and $M_{\rm P}$ is the Planck scale, which mediate $\chi$ decay. In this specific case, $\chi\to\nu\bar{\nu}$ and $\tau_{\chi\to\nu\bar{\nu}} \sim M_{\rm P}^2/(m_\chi v^2) = 1\,{\rm year}\times(1\,{\rm GeV}/m_{\chi})$. Thus, for $\chi$ to be cosmologically long lived and qualify as a dark matter candidate, the coefficient of this dimension-five operator must be engineered to be small enough, unless $\chi$ were as light as the neutrinos. There are, of course, ways to circumvent these constraints. One may, for example, consider scenarios where $U(1)_{B-L}$  is gauged and spontaneously broken to a discrete subgroup which allows $\chi$ to be stable. This would be the case if the symmetry breaking vevs only violated {\BmL} by two units~\cite{Aulakh:2000sn,Ma:2015xla,Heeck:2015qra}.

On the other hand, our results concerning $\phi$ --  a {\LeNCS} with charge two -- are mostly insensitive to $B-L$ quantum gravity effects, i.e., to Planck-suppressed higher-dimensional operators that violate $B-L$. For example, the Planck-suppressed Weinberg operator $(LH)^2/M_{\rm P}$ would give the left-handed neutrinos a very small Majorana mass and render the neutrinos pseudo-Dirac fermions. It would not, however, modify any of the results discussed in Secs.~\ref{sec:limits} and \ref{sec:oscillation}.\footnote{Things could be even safer if the model is were further supersymmetrized~\cite{Ghosh:2010hy}.} In this case, the main structure of the framework discussed throughout this work ({\BmL} as now an approximate global symmetry) can be maintained.

\section*{ACKNOWLEDGEMENTS}
We would like to acknowledge helpful discussions with Zackaria Chacko, Pilar Coloma, and Mark Wise.
JMB is supported by Department of Energy Grant No.~\#de-sc0018327 and acknowledges the support of the Colegio de F\'isica Fundamental e Interdiciplinaria de las Am\'ericas (COFI) Fellowship Program.  The work of AdG, KJK, and YZ is supported in part by Department of Energy Grant No.~\#de-sc0010143. KJK thanks the Fermilab Neutrino Physics Center for support during the completion of this work.
This manuscript has been authored by Fermi Research Alliance, LLC under Contract No. DE-AC02-07CH11359 with the U.S. Department of Energy, Office of Science, Office of High Energy Physics. The United States Government retains and the publisher, by accepting the article for publication, acknowledges that the United States Government retains a non-exclusive, paid-up, irrevocable, world-wide license to publish or reproduce the published form of this manuscript, or allow others to do so, for United States Government purposes.

%%%%%%%%
%   Appendix  %
%%%%%%%%
 
\appendix
\section{Dimension-Eight Operators with {\LeNCS} of $B-L$ Charge Two}
\label{app:dim8}

In this appendix, we present a list of dimension-eight operators that contain exactly one {\LeNCS} field as well as any number of right-handed neutrinos. To generate this list, we used the Hilbert series method described in Refs.~\cite{Henning:2015alf,Henning:2017fpj}, where the SM Hilbert series has been extended such that $U(1)_{B-L}$ is an exact symmetry of the Lagrangian.

Models of new physics generally require some finagling to produce effective operators that contain field-strength tensors; since including these in the Hilbert series significantly lengthens computation time, we have ignored them altogether. While the inclusion of the (covariant) derivative operator is a necessary ingredient in the evaluation of the Hilbert series, we do not report operators containing these in our list. Operators that are products of lower-dimension operators -- having the form $|H|^2 \times (\text{dimension-six operator})$ or $ \overline{\nu}^c \overline{\nu}^c \phi \times (\text{dimension-four operator})$ -- are uninteresting for our purposes here, and so will be ignored. 

We are interested in the weak index structure of the relevant operators, as this determines whether left-handed neutrinos will experience the interaction, but we do not report their Lorentz and color structures (except for one operator as an example, below). We do not concern ourselves with the number of independent flavor components an operator possesses; we will, however, flag operators that vanish for one generation of fermion, as these are necessarily antisymmetric in the flavor indices of at least one pair of fermions, a fact that may be relevant for phenomenology.

We find that there are 23 unique field content arrangements that yield 24 different weak-index structures, which we tabulate in Table \ref{table:dim8ops}.  Parentheses denote pairs of weak-doublet fields whose indices are to be contracted with either $\varepsilon_{ij}$ or $\delta_i^j$, as appropriate (e.g., $(L H) \to \varepsilon_{ij} L^i H^j$, while $(L H^\dagger) \to \delta_i^j L^i (H^\dagger)_j$), where $i, \, j$ ($=1,2$) are weak indices. Operators shown in green only exist for multiple generations of fermions. Asterisks denote operators with more than one possible contraction of their Lorentz indices; note that the asterisk on operator 1 is green. The last four operators are ostensibly of the form $ \overline{\nu}^c \overline{\nu}^c \phi \times (\text{dimension-4 operator})$ but, for multiple generations of fermions, it is possible to contract Lorentz indices such that these operators cannot factorize in this way. The operators in Table \ref{table:dim8ops} can all be written as $\phi$ times a dimension-seven operator that violates $U(1)_{B-L}$ by two units, consistent with the findings of Refs.~\cite{Rao:1983sd,deGouvea:2014lva,Kobach:2016ami}.

\begin{table}[!t]
\caption{List of dimension-eight operators with exactly one {\LeNCS} field and any number of right-handed neutrinos. Operators shown in green only exist for multiple generations of fermions. Asterisks denote operators with multiple possible Lorentz contractions of the same fields. Note the green asterisk on operator 1; one of its Lorentz-index configurations vanishes for one generation of fermions. The third column tabulates physical phenomena that can arise from each operator.}
\begin{center}
\begin{tabular}{c||c|c}\hline
Number & Operator & Associated Phenomena \\ \hline \hline
1$^{\tcg*}$ & $ e^c (L L) (L H) \phi $ & $\overline{\nu} e^{\pm} \to \nu e^\pm \phi$; $\ell \to \ell^\prime \nu \nu \phi$ \\ \hline
2$^*$ & $ d^c (Q L) (L H) \phi $ & $\nu p \to \ell^+ n \phi^*$; quark/meson decays \\ \hline
3 & $ \overline{u}^c (L \overline{Q}) (L H) \phi $ & $\nu p \to \ell^+ n \phi^*$, quark/meson decays \\ \hline
4 & $ \overline{\nu}^c (L \overline{L}) (L H) \phi $ & $\ell \to \ell^\prime \nu \nu \phi$; $\nu \nu \to \nu \overline{\nu} \phi^*$; C$\nu$B \\ \hline
5a & $ \overline{\nu}^c (Q \overline{Q}) (LH) \phi $ & $\overline{\nu} N \to \nu N^{(\prime)} \phi$; quark/meson decays \\ \hline
5b & $ \overline{\nu}^c (L \overline{Q}) (Q H) \phi $ & $\overline{\nu} N \to \nu N^{(\prime)} \phi$;  $\ell \to M \nu \nu \phi$; quark/meson decays \\ \hline
6 & $ d^c (L \overline{Q}) (\overline{Q} H) \phi $ & $n \to \nu \phi$; $p \to \nu \pi^+ \phi$; $\tau^- \to n \pi^- \phi^*$ \\ \hline
7 & $ \overline{\nu}^c (\overline{Q}\overline{Q}) (\overline{Q} H) \phi $ & $n \to \nu \phi$; $p \to \nu \pi^+ \phi$ \\ \hline
8 & $ \overline{\nu}^c (\overline{Q}\overline{Q}) (\overline{Q} H) \phi $ & $n \to \nu \phi$; $p \to \nu \pi^+ \phi$ \\ \hline
9$^{*}$ & $ \overline{u}^c \overline{e}^c \overline{\nu}^c (\overline{Q} H) \phi $ & $\nu p \to \ell^+ n \phi^*$; $\ell \to M \nu \nu \phi$; quark/meson decays \\ \hline
10 & $ u^c d^c d^c (L H) \phi $ & $n \to \nu \phi $; $p \to \nu \pi^+ \phi$ \\ \hline
11 & $ \overline{u}^c d^c \overline{e}^c  (L H) \phi $ & $\nu p \to \ell^+ n \phi^*$; quark/meson decays \\ \hline
12 & $ d^c \overline{d}^c \overline{\nu}^c  (L H) \phi $ & $\overline{\nu} N \to \nu N^{(\prime)} \phi$; $b$, $s$, meson decays \\ \hline
13 & $ u^c \overline{u}^c \overline{\nu}^c  (L H) \phi $ & $\overline{\nu} N \to \nu N^{(\prime)} \phi$; $t$, $c$, meson decays \\ \hline
14 & $ e^c \overline{e}^c \overline{\nu}^c  (L H) \phi $ & $\overline{\nu} e^{\pm} \to \nu e^\pm \phi$; $\ell \to \ell^\prime \nu \nu \phi$ \\ \hline
15 & $ d^c \overline{e}^c \overline{\nu}^c (Q H) \phi $ & $\nu p \to \ell^+ n \phi^*$; $\ell \to M \nu \nu \phi$; quark/meson decays \\ \hline
16 & $ u^c d^c \overline{\nu}^c (\overline{Q} H) \phi $ & $n \to \nu \phi$; $p \to \nu \pi^+ \phi$ \\ \hline
\tcg{17} & $ \tcg{d^c d^c \overline{\nu}^c ( \overline{Q} H^\dagger ) \phi} $ & \tcg{$n \to \nu K^0 \phi$; $p \to \nu K^+ \phi$} \\ \hline
\tcg{18} & $ \tcg{d^c d^c \overline{e}^c (\overline{Q} H) \phi} $ & \tcg{$n \to e^- K^+ \phi$; $\tau^- \to n K^- \phi^*$} \\ \hline
\tcg{19} & $ \tcg{d^c d^c d^c (L H^\dagger) \phi} $ & \tcg{$n \to e^- K^+ \phi$; $\tau^- \to n K^- \phi^*$} \\ \hline
\tcg{20} & $ \tcg{\overline{\nu}^c \overline{\nu}^c \overline{e}^c (\overline{L} H) \phi} $ & \tcg{$\overline{\nu} e^{\pm} \to \nu e^\pm \phi$; $\ell \to \ell^\prime \nu \nu \phi$} \\ \hline
\tcg{21} & $ \tcg{\overline{\nu}^c \overline{\nu}^c \overline{d}^c (\overline{Q} H) \phi} $ & \tcg{$\overline{\nu} N \to \nu N^{(\prime)} \phi$; $b$, $s$, meson decays} \\ \hline
\tcg{22} & $ \tcg{\overline{\nu}^c \overline{\nu}^c \overline{u}^c (\overline{Q} H^\dagger) \phi} $ & \tcg{$\overline{\nu} N \to \nu N^{(\prime)} \phi$; $t$, $c$, meson decays} \\ \hline
\tcg{23} & $ \tcg{\overline{\nu}^c \overline{\nu}^c \overline{\nu}^c (\overline{L} H^\dagger) \phi} $ & \tcg{$\nu \nu \to \nu \overline{\nu} \phi^*$; C$\nu$B} \\ \hline
\end{tabular}
\end{center}
\label{table:dim8ops}
\end{table}

The third column of Table~\ref{table:dim8ops} tabulates physical phenomena that may arise from each operator. These include:
\begin{itemize}
\item Wrong-sign lepton production in neutrino-nucleon scattering ($\nu p \to \ell^+ n \phi^*$), as discussed in the main body of this work.
\item Apparent lepton-number-violating (semi)leptonic decays of $\mu$ and $\tau$ ($\ell \to \ell^\prime \nu \nu \phi$ and $\ell \to M \nu \nu \phi$, where $\ell^{(\prime)}$ is a charged lepton and $M$ is a charged meson).
\item Apparent lepton-number-violating decays of heavy quarks and hadrons.
\item Apparent baryon-number-violating decays ($n \to \nu \phi$, $p \to \nu \pi^+ \phi$,  $\tau^- \to n \pi^- \phi^*$, etc.).
\item Neutrino-antineutrino conversion in scattering ($\overline{\nu} e^{\pm} \to \nu e^\pm \phi$ and $\overline{\nu} N \to \nu N^{(\prime)} \phi$). 
\item Neutrino self-interactions ($\nu \nu \to \nu \overline{\nu} \phi^*$) that may modify, e.g., the cosmic neutrino background (C$\nu$B).
\end{itemize}
It is beyond the scope of this work to analyze the contributions of the operators in Table~\ref{table:dim8ops} to these phenomena; we merely intend to highlight the connection between these abstract-looking operators and processes that may occur in the natural world.

As previously stated, we will not delve into the Lorentz and color structures of each operator; to do so would be cumbersome and unilluminating. However, to provide more insight into the complete form of the operators, we present, as an example, the full index structure(s) of operator 1: 
\begin{eqnarray}
e^c (L L) (L H) \phi & \to & \varepsilon_{\alpha\gamma} \varepsilon_{\beta\delta} (\varepsilon_{ij} L^{i\alpha} L^{j \beta}) (\varepsilon_{kl} L^{k\gamma} H^l) (e^c)^\delta \phi,  
\\ 
&& {\rm or} \nonumber \\
& \to & \tcg{\varepsilon_{\alpha\beta} \varepsilon_{\gamma\delta} (\varepsilon_{ij} L^{i\alpha} L^{j \beta}) (\varepsilon_{kl} L^{k\gamma} H^l) (e^c)^\delta \phi}. 
\end{eqnarray}
These operators have been written using two-component spinor notation; the left-handed lepton doublet and left-handed positron fields ($L$ and $e^c$, respectively) have spinor indices $\alpha, \, \beta, \, \gamma, \, \delta$ ($=1, \, 2$). Flavor indices have been suppressed for clarity. In this form, it is clear why the second of these vanishes for one generation of fermions: because the first two lepton doublets are antisymmetric in their Lorentz and weak indices, and because fermion fields anticommute, the operator vanishes unless these fields are antisymmetric in their flavor indices.

\bibliographystyle{apsrev-title}
\bibliography{lencs_bib}{}

\end{document}